\documentclass[12pt,preprint]{aastex}

\newcommand{\etal }{{et al.} }
\newcommand{\msun}{\thinspace M_\odot} 
\newcommand{\mjup}{\thinspace M_{\rm Jup}} 
\newcommand{\lsun}{\thinspace L_\odot} 
\newcommand{\rsun}{\thinspace R_\odot} 
\newcommand{\vect}[1]{\mbox{\boldmath$#1$}}
\def\lesssim{\mathrel{\hbox{\rlap{\hbox{\lower4pt\hbox{$\sim$}}}\hbox{$<$}}}}
\def\gtrsim{\mathrel{\hbox{\rlap{\hbox{\lower4pt\hbox{$\sim$}}}\hbox{$>$}}}}
\newcommand{\cm}{\,{\rm cm}^{-3} } 
 
\newcommand{\nc}{n_{\rm c} } 
\newcommand{\rcri}{R_{\rm c} }
\newcommand{\mdot}{M_\odot\,{\rm yr}^{-1} }
\newcommand{\tc}{t_{\rm c}}

\newcommand{\dfrac}[2]{{\displaystyle \frac{#1}{#2}} }

\shorttitle{Mass Accretion Rate}
\shortauthors{Machida \etal 2010}

\begin{document}

\title{Recurrent Planet Formation and Intermittent Protostellar Outflows Induced by Episodic Mass Accretion}
\author{Masahiro N. Machida\altaffilmark{1}, Shu-ichiro Inutsuka\altaffilmark{2}, and Tomoaki Matsumoto\altaffilmark{3}} 
\altaffiltext{1}{National Astronomical Observatory of Japan, Mitaka, Tokyo 181-8588, Japan; masahiro.machida@nao.ac.jp}
\altaffiltext{2}{Department of Physics, Nagoya University, Furo-cho, Chikusa-ku, Nagoya, Aichi 464-8602; inutsuka@nagoya-u.jp}
\altaffiltext{3}{Faculty of Humanity and Environment, Hosei University, Fujimi, Chiyoda-ku, Tokyo 102-8160, Japan; matsu@hosei.ac.jp}

\begin{abstract}
The formation and evolution of a circumstellar disk in magnetized cloud cores is investigated from prestellar core stage until $\sim10^4$\,yr after protostar formation.
In the circumstellar disk, fragmentation first occurs due to gravitational instability in a magnetically inactive region, and substellar-mass objects appear.
The substellar-mass objects lose their orbital angular momenta by gravitational interaction with the massive circumstellar disk and finally fall onto the protostar. 
After this fall, the circumstellar disk increases its mass by mass accretion and again induces fragmentation.
The formation and falling of substellar-mass objects are repeated in the circumstellar disk until the end of the main accretion phase.
In this process, the mass of fragments remain small, because the circumstellar disk loses its mass by fragmentation and subsequent falling of fragments before it becomes very massive.
In addition, when fragments orbit near the protostar, they disturb the inner disk region and promote mass accretion onto the protostar.
The orbital motion of substellar-mass objects clearly synchronizes with the time variation of the accretion luminosity of the protostar.
Moreover, as the objects fall, the protostar shows a strong brightening for a short duration.
The intermittent protostellar outflows are also driven by the circumstellar disk whose magnetic field lines are highly tangled owing to the orbital motion of fragments.
The time-variable protostellar luminosity and intermittent outflows may be a clue for detecting planetary-mass objects in the circumstellar disk.
\end{abstract}
\keywords{accretion, accretion disks: ISM: clouds---stars: formation---stars: low-mass, brown dwarfs: planetary systems: planetary disks}

\section{Introduction}
The observation of many star-forming regions has shown that stars form in molecular cloud cores.
A star is born after a long journey through the gravitational collapse of a cloud core.
The gas begins to collapse in a dense part of the cloud (i.e., the cloud core) and increases its density and temperature as the cloud core collapses further.
When the collapsing gas reaches a sufficiently high density ($\sim10^{21}\cm$), the collapse stops and a protostar having almost a Jovian mass is born \citep{larson69}.
Then, the protostar increases its mass through mass accretion and finally evolves into a main-sequence star (or main-sequence phase).

The star formation process can be divided into two phases: the early protostellar collapse phase and later accretion phase of the star formation.
The early phase, also called the gas collapsing phase, is defined as the period before protostar formation after gas collapse is initiated in the cloud core.
Thus, during this phase, the gas continues to collapse.
On the other hand, the later phase is defined as the period before the main-sequence phase after the gas collapse stops and a protostar appears in the collapsing cloud core.
In particular, during the later phase, the period when the protostar significantly increases its mass by mass accretion is called ``the main accretion phase.''
Since the star acquires almost all its mass in this main accretion phase, the final stellar mass is determined in this phase.
Protostars during the main accretion phase are observed (or defined) as Class 0 or I objects by spectral energy distribution (SED).
The observation of star-forming regions also has shown that almost all Class 0 or I objects drive a protostellar outflow.
Because a certain fraction of mass and angular momentum are ejected by the protostellar outflows, they are closely related to the rate of mass accretion onto the protostar and final stellar mass.
Moreover, the infalling gas with an angular momentum forms the circumstellar disk during the main accretion phase.
The evolution of the circumstellar disk is related to the mass accretion and outflow rates.
In addition, the circumstellar disk is the site of planet formation.
Therefore, understanding the evolution of the circumstellar disk during the main accretion phase is essential to understanding both the star and planet formation processes.

Since the protostar and circumstellar disk during the main accretion phase are veiled by the infalling envelope, it is difficult to observe them.
Recently, however, the {\it Spitzer Space Telescope} (SST) has been unveiling this phase.
\citet{enoch09a} discovered  a massive circumstellar disk of $\sim1\msun$ comparable to a central protostar around a Class 0 object, indicating that (i) the  disk already exists in the main accretion phase, and (ii) the disk mass is significantly larger than the theoretical prediction.
\citet{evans09} observed several star forming regions and identified several Class 0 objects using the SST.
They showed that the bolometric luminosity of the objects is considerably dimmer than classical theoretical predictions, and has a dispersion over 2-3 orders of magnitude.
They pointed out that recent observations aggravate the ``luminosity problem'' (the problem that the accretion luminosity of protostars is lower than  theoretical predictions, see \citealt{kenyon90}), and concluded that non-steady accretion is inevitably required to explain the observational results.
The non-steady accretion may be attainable when the circumstellar disk is sufficiently massive and causes the gravitational instability \citep{durisen07}.

Another recent development in observation of star and planet formation is direct images of exo-planets \citep{kalas08,marois08,thalmann09}, in which planets are located at $\gtrsim 10$\,AU from the central star.
However, in the framework of the core accretion scenario \citep{hayashi85}, there is less likehood of planets forming in a region that is so remote from the central star.
Alternatively, the gravitational instability scenario \citep{cameron78} may explain the formation of such planets.
In summary, recent observations seem to indicate that the protostars have a considerably massive disk unlike what was previously believed.

Although it seems that recent observational progress is unveiling problems for the early evolution stage of star formation, we cannot directly observe the circumstellar disk and protostar in the main accretion phase because they are embedded in the dense infalling envelope.
Thus, theoretical study is necessary to understand the properties of the circumstellar disk and protostar.
In particular, multi-dimensional simulations are necessary to investigate the evolution of the circumstellar disk, protostellar outflow, and so on.

The evolution of the collapsing gas cloud core from the protostellar core stage (i.e., the gas collapsing phase) until protostar formation has been well investigated using multi-dimensional simulations \citep[e.g.,][]{bate98,tomisaka02,whitehouse06,stamatellos07,banerjee06,machida06,machida07,machida08a,machida08b,machida09b}.
On the other hand, only a few studies have focused on the main accretion phase immediately following the prestellar core stage, because it is difficult to calculate long-term evolution in the main accretion phase with sufficient spatial resolution.
In unmagnetized cloud cores, the formation and evolution of the circumstellar disk in the main accretion phase through the gas collapsing phase were investigated by \citet{walch09a,walch09b}, \citet{vorobyov10}, and \citet{machida10}.
They found that the circumstellar disk is considerably massive to induce fragmentation or the gravitational instability that is related to a non-steady accretion flow onto the protostar.

In reality, however, since molecular clouds are strongly magnetized \citep{crutcher99}, the magnetic field may play an important role in the evolution of the circumstellar disk during the main accretion phase.
\citet{vorobyov06,vorobyov07} investigated the evolution in two dimensions of the circumstellar disk in a magnetized cloud core and showed the non-steady accretion onto the central protostar.
In three dimensions, the formation and evolution of the circumstellar disk from prestellar core stage were investigated only by \cite{inutsuka09}, in which they showed fragmentation and possible planet formation in the magnetically inactive region of the circumstellar disk during the main accretion phase.
They also indicated non-steady mass accretion onto the protostar owing to the gravitational instability of the circumstellar disk.
However, this study only calculated the evolution of the circumstellar disk about $\sim1000$\,yr after protostar formation.

In this study, in a setting similar to \cite{inutsuka09}, we investigate the evolution of the circumstellar disk for $\sim10^{4}$\,yr, which is $\sim10$ times longer than the previous study. 
In addition to the model adopted by \citet{inutsuka09}, we newly calculate the evolution of the circumstellar disk formed in a relatively stable initial cloud core.
In both models, we compare the mass accretion rate, properties of protostellar outflow, and the fragmentation condition.
The structure of the paper is as follows. 
The framework of our models and the numerical method are given in \S 2. 
The numerical results are presented in \S 3. 
We discuss the fragmentation condition of the circumstellar disk and its implication for the planet formation in \S 4, and we summarize our results in \S 5.

\section{Model and Numerical Method}
\label{sec:model}
\subsection{Basic Equations}
To study the formation and evolution of a circumstellar disk in a magnetized molecular cloud core, we solve the three-dimensional resistive MHD equations, including self-gravity:
\begin{eqnarray} 
& \dfrac{\partial \rho}{\partial t}  + \nabla \cdot (\rho \vect{v}) = 0, & \\
& \rho \dfrac{\partial \vect{v}}{\partial t} 
    + \rho(\vect{v} \cdot \nabla)\vect{v} =
    - \nabla P - \dfrac{1}{4 \pi} \vect{B} \times (\nabla \times \vect{B})
    - \rho \nabla \phi, & \\ 
& \dfrac{\partial \vect{B}}{\partial t} = 
   \nabla \times (\vect{v} \times \vect{B}) + \eta_{\rm OD} \nabla^2 \vect{B}, & 
\label{eq:reg}\\
& \nabla^2 \phi = 4 \pi G \rho, &
\end{eqnarray}
where $\rho$, $\vect{v}$, $P$, $\vect{B} $, $\eta_{\rm OD}$, and $\phi$ denote density, velocity, pressure, magnetic flux density, resistivity, and gravitational potential, respectively.
In addition, we adopted the hyperbolic divergence cleaning method of \citet{dedner02} to obtain divergence-tree magnetic field ($\vect{\nabla} \cdot \vect{B}=0$).
With this method, no magnetic monopoles appear throughout calculation \citep[see also][]{machida05a,matsu07}.

To mimic the temperature evolution calculated by \citet{masunaga00}, we adopt the piece-wise polytropic equation of state \citep[see][]{vorobyov06,machida07} as\begin{equation} 
P =  c_{s,0}^2\, \rho \left[ 1+ \left(\dfrac{\rho}{\rho_c}\right)^{2/5} \right],
\label{eq:eos}
\end{equation}
 where $c_{s,0} = 190$\,m\,s$^{-1}$ and 
$ \rho_c = 3.84 \times 10^{-14} \, \rm{g} \, \cm$ ($n_c = 10^{10} \cm$). 
With equation~(\ref{eq:eos}), the gas behaves isothermally for $n\lesssim10^{10}\cm$ and adiabatically for $n\gtrsim10^{10}\cm$.
For a realistic evolution of the magnetic field in the circumstellar disk, we adopt the resistivity ($\eta_{\rm OD}$) as the fiducial value in \citet{machida07}, in which the Ohmic dissipation becomes effective for $10^{11}\cm \lesssim n \lesssim 10^{15}\cm$ (for details, see Eq.~[9] and [10], and Fig.~1 of \citealt{machida07}).

\subsection{Initial Setting and Numerical Method}
\label{sec:setting}
As the initial state, we take a spherical cloud core with a critical Bonnor--Ebert (BE) density profile, in which a uniform density ($\rho_{\rm amb}\simeq0.07\rho_c$) is adopted outside the sphere ($r > \rcri$).
For the BE density profile, we adopt the central density as $\nc =  10^{6}\cm$ and isothermal temperature as $T=10$\,K. 
For these parameters, the critical BE radius is $\rcri = 4.7\times10^3$\,AU.
The gravitational force is ignored outside the host cloud ($r>\rcri$) to mimic a stationary interstellar medium.
In addition, we prohibit the gas inflow at $r=\rcri$ to suppress mass input from the interstellar medium into the gravitationally collapsing cloud core.
Thus, only the gas inside $r < \rcri$ collapses to form the circumstellar disk and protostar.
Note that although the protostellar outflow driven by the circumstellar disk propagates into the region of $r > \rcri$ and disturbs the interstellar medium over time, we can safely calculate the mass accretion process onto the circumstellar disk because the outflow propagating into the interstellar medium does not affect the inner cloud core region  ($r < \rcri$).
In this study, we call this initial spherical cloud with Bonnor--Ebert density profile `the cloud core' that has a radius of $\rcri$ and mass of $M_{\rm cl}$.

Since the critical BE sphere is in an equilibrium state, we increase the density by a factor of $f$ to promote the contraction, where $f$ is the density enhancement factor that represents the stability of the initial cloud core.
An initial cloud core with larger $f$ is more unstable against gravity.
In general, the stability of the cloud core is represented by a parameter $\alpha_0$ ($\equiv E_{\rm th}/E_{\rm grav}$) that is the ratio of thermal ($E_{\rm th}$) to gravitational ($E_{\rm grav}$) energy.
As shown in \citet{matsu03}, when the BE density profile is adopted, the density enhancement factor is related to the parameter $\alpha_0$ as 
\begin{equation}
\alpha_0 = \dfrac{0.84}{f}.
\label{eq:alpha}
\end{equation}
We constructed two models with different $f$ ($=1.68$ and $2.8$).
In this paper, we call the model having $f=1.68$ ($\alpha=0.5$) ``model A05,'' and the model having $f=2.8$ ($\alpha=0.3$) ``model A03.''
At the initial state, model A03 is more unstable than model A05.

In both models, the initial cloud core rotates rigidly around the $z$-axis in the region of  $r< \rcri$, while the uniform magnetic field parallel to the $z$-axis (or rotation axis) is adopted in the whole computational domain.
In addition, we adopt that the angular velocity decreases in proportional to $\propto {\rm exp} (-r^2)$ outside the host cloud ($r>\rcri$).
Both models have the same angular velocity with $\Omega_0 = 1.1 \times 10^{-13}$\,s$^{-1}$ and the same magnetic field, $B_0 =37 $$\mu$G.
The mass inside $r < \rcri$ for each model is $0.8\msun$ (model A05) and $1.3\msun$ (model A03).
Model names and initial values are summarized in Table~1.
The initially magnetized, rotating cloud core is also specified by the non-dimensional parameters $\beta_0$ and $\gamma_0$, where $\beta_0$ ($\equiv E_{\rm rot}/\vert E_{\rm grav} \vert$) and $\gamma_0$ ($\equiv E_{\rm mag}/\vert E_{\rm grav} \vert$) is the ratio of the rotational and magnetic energy to the gravitational energy inside the initial cloud core.
Model A03 has $\beta_0=5\times10^{-3}$ and $\gamma_0=0.05$, while model A05 $\beta_0=6\times10^{-3}$ and $\gamma_0=0.07$.
In addition, the mass-to-flux ratio $M/\Phi$ is also used to specify the initial cloud, 
where $M$ and $\Phi$ are the mass and magnetic flux of the initial cloud core.
There exists a critical value of $M/\Phi$ below which a cloud is supported against the gravity by the magnetic field.
For a cloud with uniform density,  \citet{mouschovias76} derived a critical mass-to-flux ratio
\begin{equation}
\left(\dfrac{M}{\Phi}\right)_{\rm cri} = \dfrac{\zeta}{3\pi}\left(\dfrac{5}{G}\right)^{1/2},
\label{eq:mag2}
\end{equation}
where the constant $\zeta=0.53$.
We define the mass-to-flux ratio normalized by the critical value as $\lambda$.
In our setting, model A03 has $\lambda=9$, while model A05 has $\lambda=5.6$.

We add $m=2$-mode non-axisymmetric density perturbation to the initial cloud core.
Then, the density profile of the cloud core is described as  
\begin{eqnarray}
\rho(r) = \left\{
\begin{array}{ll}
\rho_{\rm BE}(r) \, (1+\delta_\rho)\,f & \mbox{for} \; \; r < R_{c}, \\
\rho_{\rm BE}(R_c)\, (1+\delta_\rho)\,f & \mbox{for}\; \;  r \ge R_{c}, \\
\end{array}
\right. 
\end{eqnarray}
where $\rho_{\rm BE}(r)$ is the density distribution of the critical 
BE sphere, and $\delta_\rho$ is the axisymmetric density perturbation. 
For the $m=2$-mode, we chose
\begin{equation}
\delta_\rho = A_{\phi} (r/R_{\rm c})^2\, {\rm cos}\, 2\phi, 
\label{eq:dens-pert}
\end{equation}
where $A_{\phi}$ (=0.01) represents the amplitude of the perturbation.
The radial dependence is chosen so that the density perturbation remains regular at the origin ($r = 0$) at one time-step after the initial stage.
This perturbation ensures that the center of gravity is always located at the origin.

In the collapsing cloud core, we assume the protostar formation to occur when the number density exceeds $n > 10^{13}\cm$ at the center.
To model the protostar, we adopt a sink around the center of the computational domain.
In the region $r < r_{\rm sink} = 1\,$AU, gas having a number density of $n > 10^{13}\cm$ is removed from the computational domain and added to the protostar as  gravity in each timestep \citep[for details, see][]{machida09a,machida10}.
This treatment of the sink makes it possible to calculate the evolution of the collapsing cloud core and circumstellar disk for a longer duration.
In addition, inside the sink, the magnetic flux is removed by Ohmic dissipation, because such a region has the magnetic Reynolds $Re$ number exceeding unity $Re>1$ \citep[for details, see][]{machida07}.

To calculate on a large spatial scale, the nested grid method is adopted \citep[for details, see ][]{machida05a,machida05b}. 
Each level of a rectangular grid has the same number of cells ($ 128 \times 128 \times 32 $).
The calculation is first performed with five grid levels ($l=1-5$).
The box size of the coarsest grid $l=1$ is chosen to be $2^5 \rcri$.
Thus, grid of $l=1$ has a box size of $\sim 1.5\times 10^5$\,AU.
A new finer grid is generated before the Jeans condition is violated \citep{truelove97}.
The maximum level of grids is $l_{\rm max} = 12$ that has a box size of 74\,AU and a cell width of 0.58\,AU.

We adopted a fixed boundary condition on the outermost grid boundary.
The uniform density ($\rho_{\rm amb}$) and magnetic field ($B_x=0$, $B_y=0$, $B_z=B_0$) and zero fluid velocity ($v_x = v_y = v_z = 0$) are imposed on $l=1$ grid boundary at each timestep.
Such boundary condition hardly affects the evolution of the cloud core, because the gravitationally collapsing cloud core with a radius of $4.7\times10^3$\,AU ($r=\rcri$) is embedded in a large simulation box with a size of $\sim 3\times 10^5$\,AU, inside which the static interstellar medium is fulfilled in the region of $r>\rcri$.
In our setting, the Alfv$\acute{\rm e}$n speed in the interstellar medium ($r>r_{\rm BE}$) for model A03 is $v_{\rm A}=0.15$\,km\,s$^{-1}$.
Thus, by reaching the Alfv$\acute{\rm e}$n wave to the computational boundary from the center of the cloud core, it takes $4.7\times10^6$\,yr.
Since we stopped our calculation in $\sim10^4$\,yr after the calculation begins, the Alfv$\acute{\rm e}$n wave generated at the center of the cloud core (or the computational boundary) never reaches the computational boundary (or the center of the cloud core).

\section{Results}
We calculated the formation and evolution of the circumstellar disk in magnetized collapsing cloud cores.
First, we simply outline the formation and evolution of the circumstellar disk.
Before protostar formation, the first (adiabatic) core forms in the collapsing cloud core with a scale of $\sim10$\,AU \citep{larson69,masunaga00,saigo06}.
At its formation, the first core has a thick disk-like structure and is mainly supported by the thermal pressure gradient force.
After protostar formation, the first core becomes thin because the centrifugal force dominates the thermal pressure gradient force \citep{walch09a,machida10}.
During the main accretion phase, the first core grows and extends to a large extent ($\sim100-1000$\,AU).
\citet{inutsuka09} and \cite{machida10} pointed out that the circumstellar disk  originates from the first core, and is formed before protostar formation \citep[see also][]{bate98,bate10}.
Their calculations showed that the first core gradually grows to become the circumstellar disk in the main accretion phase.
They also pointed out that, reflecting the thermal history of the collapsing cloud core,  the circumstellar disk is inevitably more massive than the protostar in the early part of the main accretion phase.
The formation of the first core (or the circumstellar disk) before protostar formation has been investigated in many past studies \citep[see review by][]{bodenheimer00,goodwin07}.
Thus, although we began the calculation from the prestellar core stage, we mainly focus on the evolution of the circumstellar disk in the main accretion phase in this paper.

\subsection{Recurrent Fragmentation and Planet Formation}
As described in \S\ref{sec:model}, we constructed two models (models A03 and A05) with different initial stabilities (i.e., different $\alpha_0$).
In the two models,  the protostar forms $t \simeq 2.50\times10^4$\,yr (model A03) and $t \simeq 5.48\times10^4$\,yr (model A05), respectively, after the calculation begins.
Thus, in the collapsing cloud core, the protostar for model A03 forms for a shorter duration than that for model A05.
This is because the initial cloud core for model A03 is more unstable than that for model A05 and  begins to collapse according to the self-similar solution \citep{larson69} in a shorter time after the calculation begins \citep{machida08a}.
In both models, we calculated the evolution of the circumstellar disk for $\sim 10^4$\,yr after protostar formation.

\begin{figure}
\includegraphics[width=150mm]{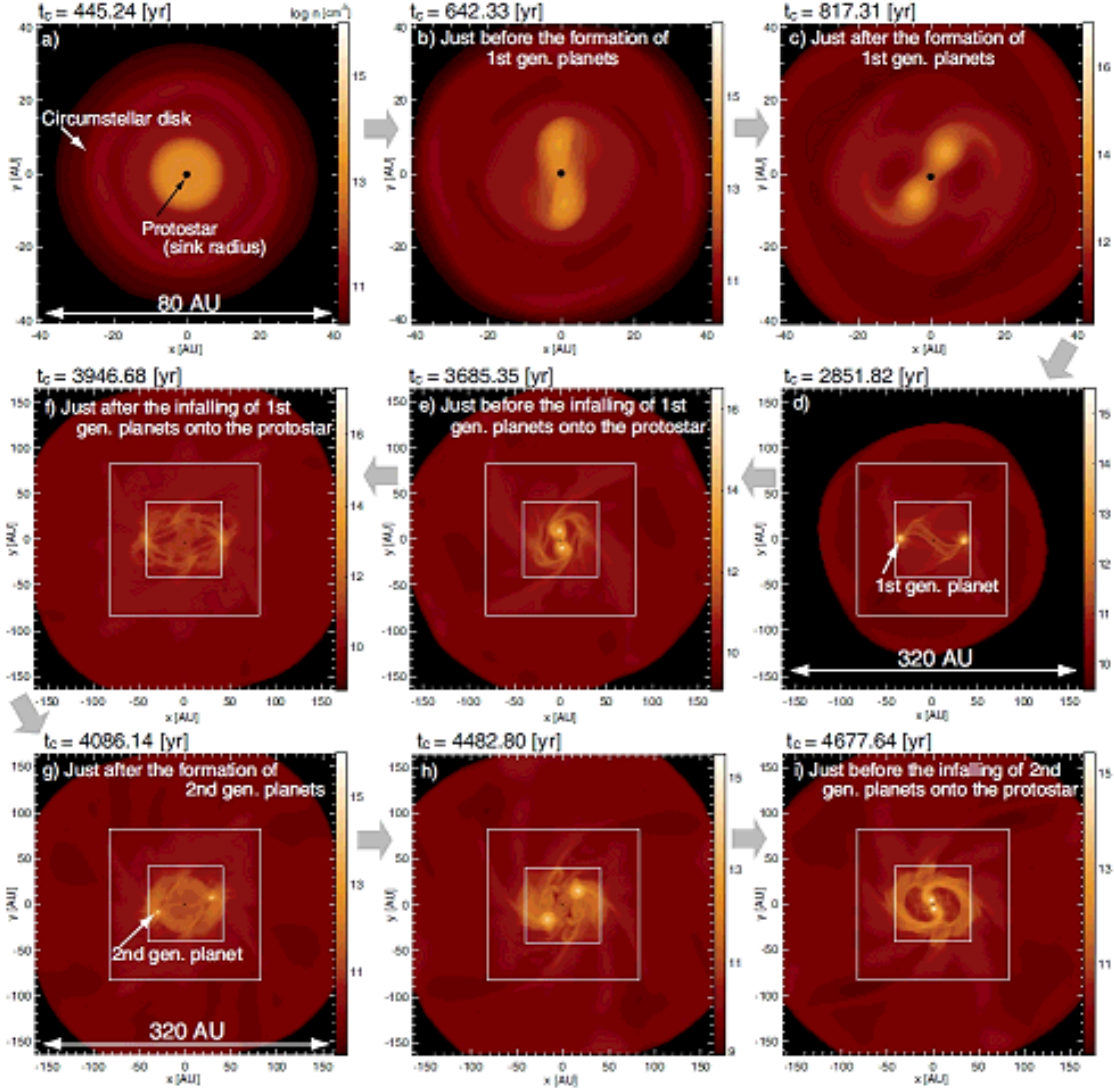}
\caption{
Time sequence images of the circumstellar disk after protostar formation for model A03.
The density distribution on the equatorial plane around the protostar is plotted in each panel.
The size of the black circle at the center in each panel corresponds to the sink radius.
The box size of the upper three panels is $\sim80$\,AU, while that of the middle and lower panels is $\sim320$\,AU.
The white squares in the middle and lower panels denote the outer boundary of the subgrid.
The time elapsed after the protostar formation $\tc$ is plotted on the upper side of each panel.
}
\label{fig:1}
\end{figure}

First, we describe the evolution of the circumstellar disk for model A03 that shows the recurrent fragmentation and  formation of substellar-mass object in the main accretion phase.
Figure~\ref{fig:1} shows the evolution of the circumstellar disk after protostar formation for model A03, in which the density distributions around the protostar on the equatorial plane are plotted.
The box sizes in the upper panels are $\sim80$\,AU, while those in the middle and lower panels are $\sim320$\,AU.
Figure~\ref{fig:1}{\it a} shows the structure of the circumstellar disk $\tc \simeq 450$\,yr after  protostar formation, in which the red region ($n\gtrsim10^{11}\cm$) corresponds to the circumstellar disk.
Note that we describe the elapsed time after protostar formation with a notation of `$\tc$,' which differs from the elapsed time `$t$' after the calculation begins (or the cloud begins to collapse).
The size of the circumstellar disk increases with time and extends up to $\sim$285\,AU by the end of the calculation.
In the circumstellar disk, fragmentation occurs due to gravitational instability $\tc \simeq 630$\,yr after protostar formation \citep[for details, see][]{inutsuka09,machida10}.
Two ambiguous clumps appear in Figure~\ref{fig:1}{\it b} and {\it c}, while two clear fragments are seen in Figure~\ref{fig:1}{\it d}.
Fragmentation occurs $\sim 6.8$\,AU away from the protostar.
Then, fragments orbit around the protostar with an orbital separation of $\sim2-38$\,AU (Figs.~\ref{fig:1}{\it c}-{\it e}) and finally fall onto the protostar at $\tc \simeq 3760$\,yr (Figs.~\ref{fig:1}{\it e} and {\it f}).
In this section, we call the fragment appeared in the circumstellar disk the planet.
Note that in reality, since some fragments that appeared in this calculation exceeded the deuterium-burning limit (about 13 Jupiter mass), they are not planets in the complete sense.
Note also that we define, for convenience, the objects born in the circumstellar disk as planets to characterize them as a whole: we discuss their mass in \S\ref{sec:mass}.

\begin{figure}
\includegraphics[width=150mm]{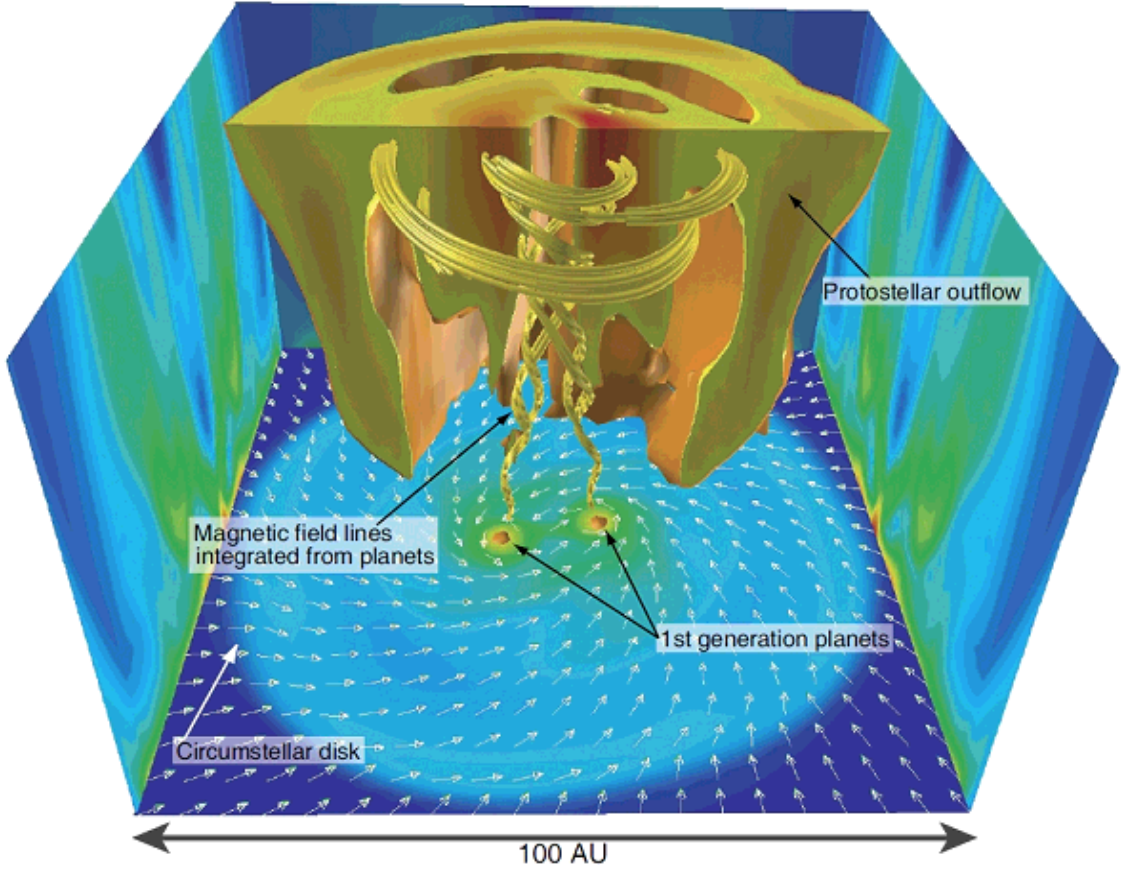}
\caption{
Configuration of protostellar outflow at $\tc = 843$\,yr is shown by yellow volume, in which color indicates outflow speed.
The density distributions (colors) are projected on each wall surface.
The velocity vectors (arrows) on the equatorial plane are plotted on bottom surface. 
The magnetic field lines integrated from each planet are plotted by yellow lines.
The box size is 100\,AU.
}
\label{fig:2}
\end{figure}

Figure~\ref{fig:2} shows the structure of the circumstellar disk and protostellar outflow at almost the same epoch as Figure~\ref{fig:1}{\it c} in three-dimensions.
The outflow driven by the first core in the collapsing cloud core has been studied by many authors \citep[e.g.,][]{matsu04,machida04,machida05a, hennebelle08a}.
Even in the main accretion phase (i.e., even after protostar formation), the circumstellar disk that originates from the first core continues to drive outflow \citep{tomisaka02,banerjee06,machida07,machida09a}.
In this model, the outflow driven by the circumstellar disk extends up to $\sim$516\,AU by the end of the calculation.
As seen in Figure~\ref{fig:2}, the outflow is driven only by the outer disk region, while the outflow is not driven by the inner region of the circumstellar disk where fragmentation occurs and planets appear.
This is because the inner disk region has a density of  $n \gtrsim 10^{11}-10^{12}\cm$ and the magnetic field is dissipated by Ohmic dissipation.
As shown in \citet{nakano02} and \citet{machida07}, when the number density exceeds $n\gtrsim 10^{11}-10^{12}\cm$, Ohmic dissipation becomes effective owing to the extremely low ionization degree.
Note that the ambipolar diffusion can be effective in the range of $n<10^{11}-10^{12} \cm$ (see, \S\ref{sec:ambipolar}).
Fragmentation is suppressed by the magnetic field because magnetic effects such as magnetic braking and outflows effectively transfer the angular momentum that promotes fragmentation \citep{machida05b,machida08a,hennebelle08b}.
Thus, a weaker field promotes fragmentation in the circumstellar disk.
As a result, the inner disk region has a considerable weak magnetic field and fragmentation occurs without outflow, while the outer disk region has a strong field and no fragmentation occurs with outflow.

In Figure~\ref{fig:2}, yellow lines  are magnetic field lines.
In the figure,  only magnetic field lines that threaded planets are  plotted to stress the planet's spin or orbital motion.
The magnetic field lines just above each planet are strongly twisted by the spin motion of the planet.
In the midair (or on a large scale), the magnetic field lines originating from each planet are twisted together, reflecting the orbital motion of the planets.
Although we only plotted  magnetic field lines near the planet to understand the planet's motion, the magnetic field (or Lorentz force) in this region (or inner disk region) is extremely weak ($\beta_{\rm p}\gg10$, where $\beta_{\rm p}$ is the plasma beta and different from initial ratio of the rotational to the gravitational energy, $\beta_0$) because Ohmic dissipation is effective.
Thus, near the planets, the magnetic field lines are passively moved, and they hardly affect the dynamical evolution of both the planet and circumstellar disk.
However, planets perturb the outer disk region that is connected to magnetic field lines driving outflow (i.e., the magnetic field lines in the outer disk region where  Ohmic dissipation is not effective).
As a result,  a highly time-variable outflow appears in this model.
We will discuss an intermittent outflow in \S\ref{sec:outflow}.

\begin{figure}
\includegraphics[width=150mm]{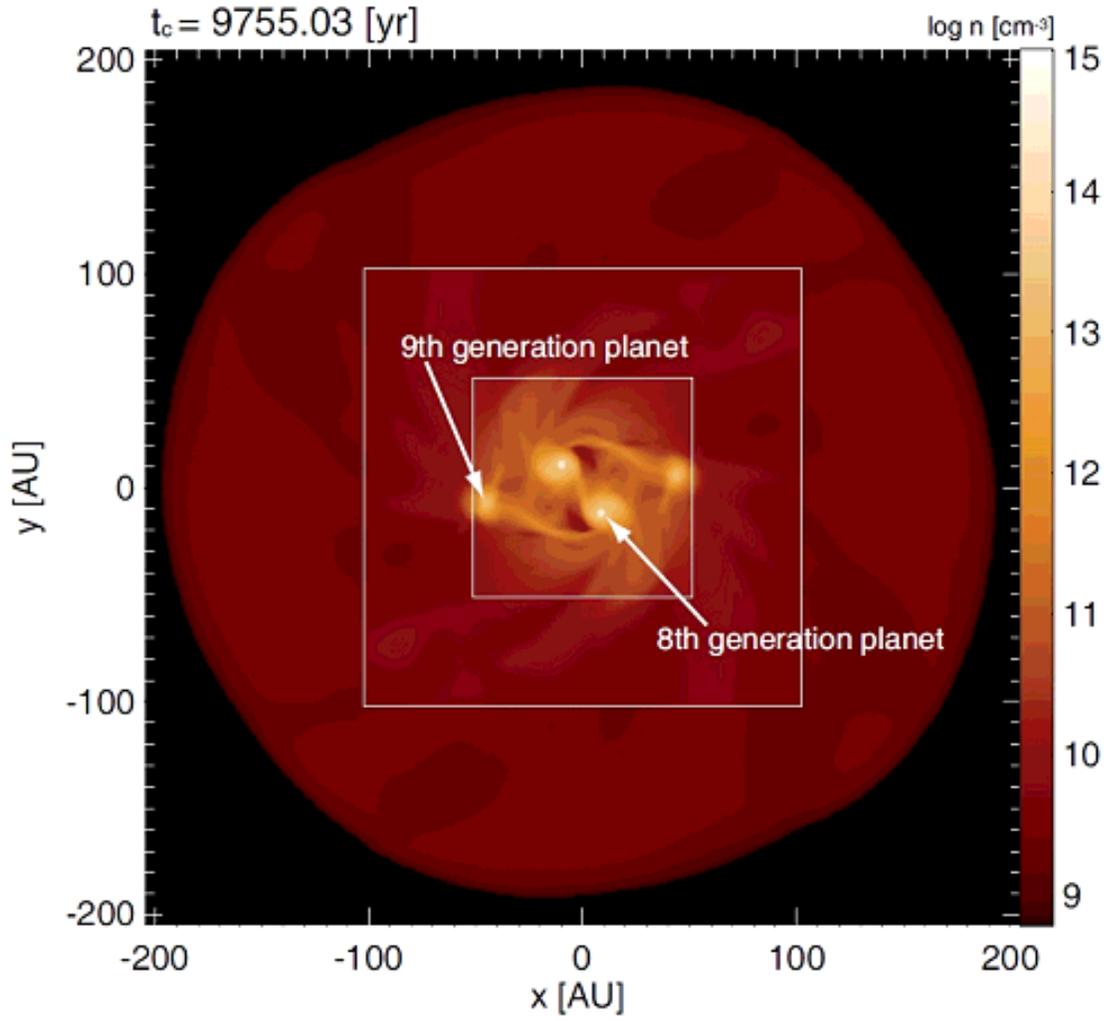}
\caption{
Density distribution of circumstellar disk on equatorial plane $\tc =9755$\,yr after protostar formation.
Eighth- and ninth-generation planets are shown.
}
\label{fig:3}
\end{figure}

Figure~\ref{fig:1}{\it e} shows the structure of the circumstellar disk just before the planets fall onto the protostar, while Figure~\ref{fig:1}{\it f} shows the disk just after the fall.
After the first-generation planets disappear, fragmentation occurs again in the circumstellar disk and the second-generation planets appear $\tc \simeq 3990$\,yr after protostar formation as shown in Figure~\ref{fig:1}{\it g}.
However, these planets also fall onto the protostar $\sim680$\,yr after their formation.
The formation and falling of the planets are repeated several times  in the circumstellar disk.
By the end of the calculation, 18 planets appeared and 16 planets fell onto the protostar.
Figure~\ref{fig:3} shows eighth- and ninth-generation planets.
The eighth-generation planets fell onto the protostar, while the ninth-generation planets survived  until the end of the calculation.
The orbital radius $r_i$, time  $t_i$ at planet formation, planet's falling epoch $t_e$, lifetime $t_e-t_i$, and mass at its formation $M_{\rm pl}$ are summarized in Table~2.

\subsection{Orbital Trajectories of Planets}
\begin{figure}
\includegraphics[width=150mm]{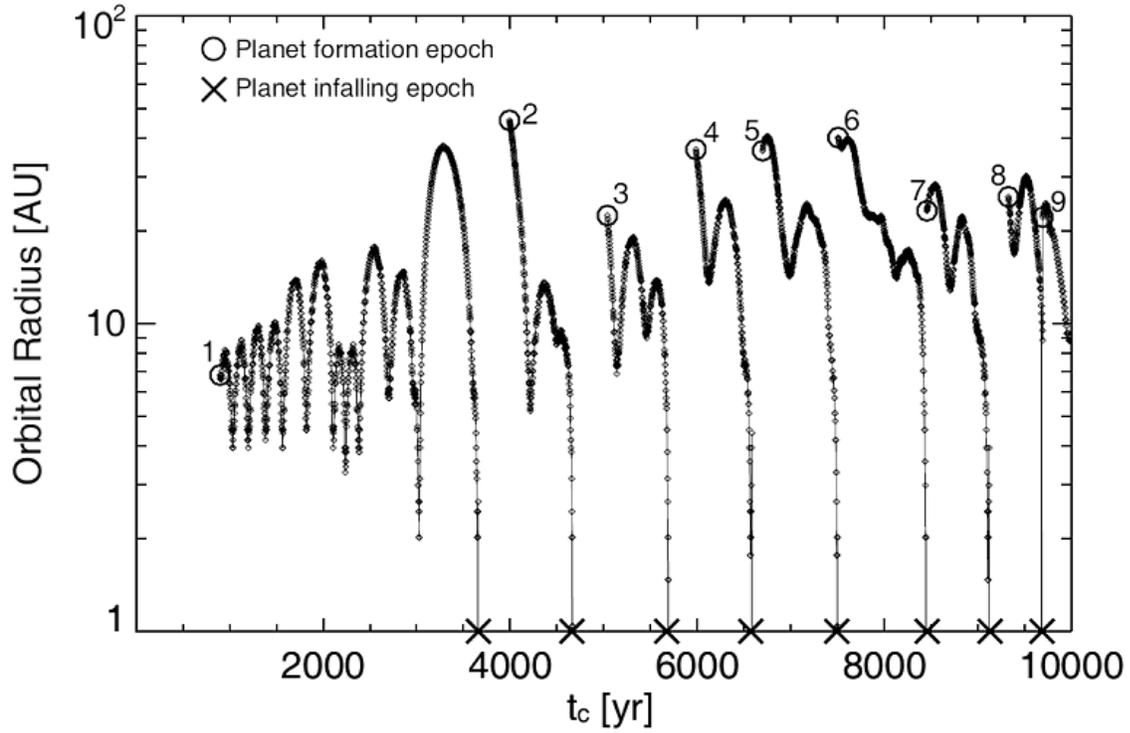}
\caption{
Orbital evolution of the most massive planet against time elapsed after protostar formation.
Symbol {\small $\bigcirc$} indicates the epoch at which fragmentation occurs and planet appears, while symbol  $\times$ is the epoch at which planet falls onto  protostar. 
Numbers 1-9 corresponds to generation of planet.
}
\label{fig:4}
\end{figure}

Figure~\ref{fig:4} shows the orbital evolution of planets against time after protostar formation $\tc$.
When multiple planets appear as seen in Figures~\ref{fig:1} and \ref{fig:3}, only the orbital radius of the most massive planet is plotted.
The figure indicates that the first-generation planets form at $\sim6.8$\,AU from the protostar at $\tc\simeq795$\,yr, and shrink their orbit and fall onto the protostar after $\sim10$ orbital rotations.
The second-generation planets appear at $\sim46$\,AU and fall onto the protostar after $\sim2-3$ orbital rotations.
This formation and falling of planets is repeated, until the ninth-generation planet is formed.

\begin{figure}
\includegraphics[width=150mm]{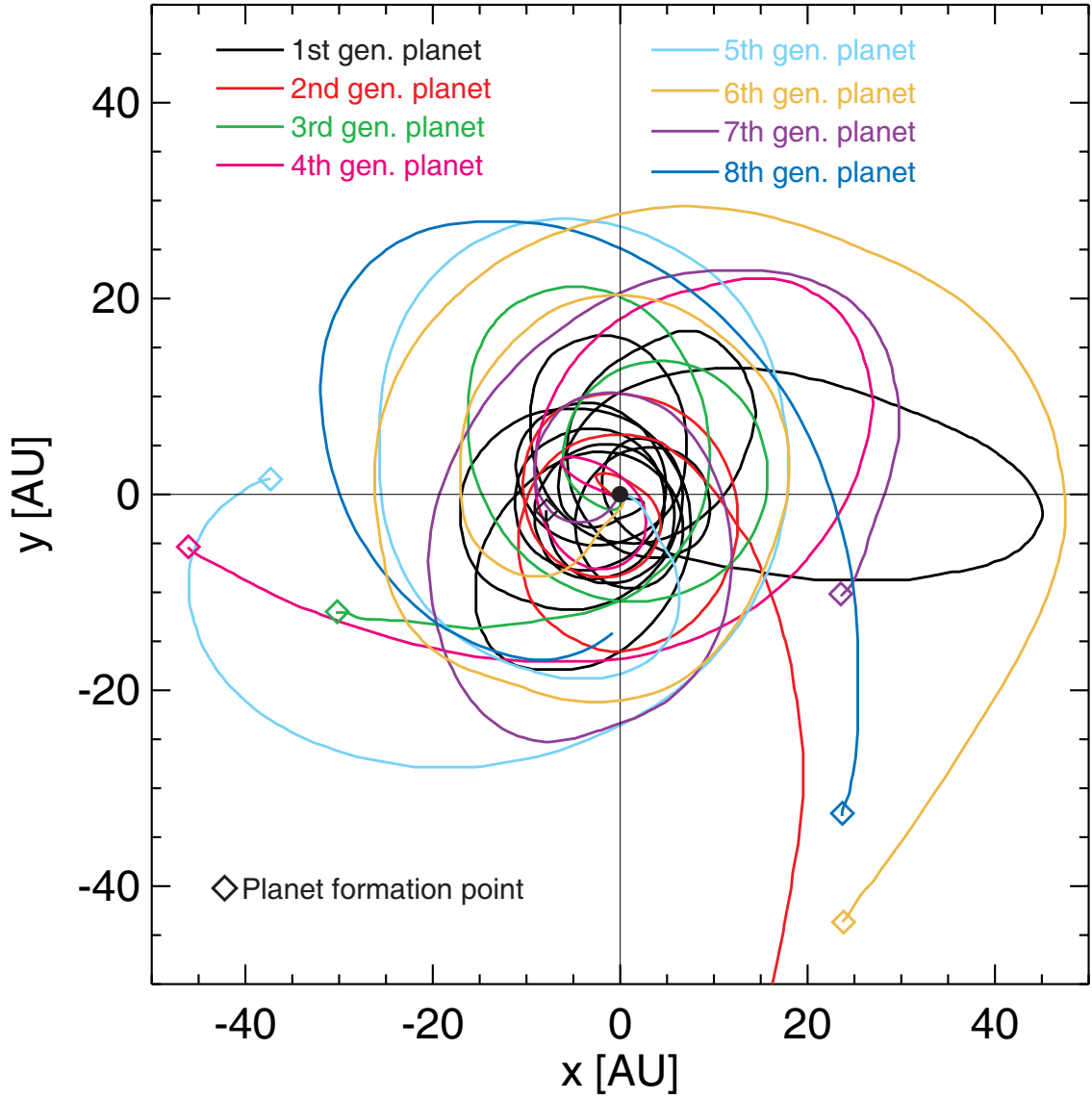}
\caption{
Trajectory for each planet just after planet formation (diamonds) until planet falls onto protostar (1st-8th-generation planets) or end of calculation (9th-generation planet).
Size of black circle at center corresponds to sink radius.
}
\label{fig:5}
\end{figure}

Figure~\ref{fig:5} shows the trajectory for planets for each generation.
The figure indicates that planets shrink their orbits with a certain amount of eccentricity.
Figures~\ref{fig:4} and \ref{fig:5} show that planets are born at $\sim6-50$\,AU  from the protostar and orbit for $\sim500-1000$\,yr (or $2-10$ orbital rotations) around the protostar.
In addition, these figures show that the first-generation planets are formed at a relatively small orbital radius ($\sim 6.5$\,AU) and orbit for $\sim10$ times around the protostar, while other generations of planets are formed at a larger orbital radii ($\sim20-50$\,AU) and orbits only for $2-3$ times.
This is because the circumstellar disk increases its size and mass with time.
As shown in Figure~\ref{fig:1}, the circumstellar disk has a size of $\sim20-30$\,AU when the first-generation planet appears, while the disk size reaches $\sim200$\,AU just before the second-generation planet appears.
As a result, the magnetically inactive zone also expands as the disk mass and radius increase, and fragmentation and planet formation can occur in such an outer disk region.
The mass evolution of the circumstellar disk is shown in \S\ref{sec:mass1}, and the fragmentation condition is discussed in \S\ref{sec:property}.

\subsection{Non-Fragmentation Model}
\begin{figure}
\includegraphics[width=150mm]{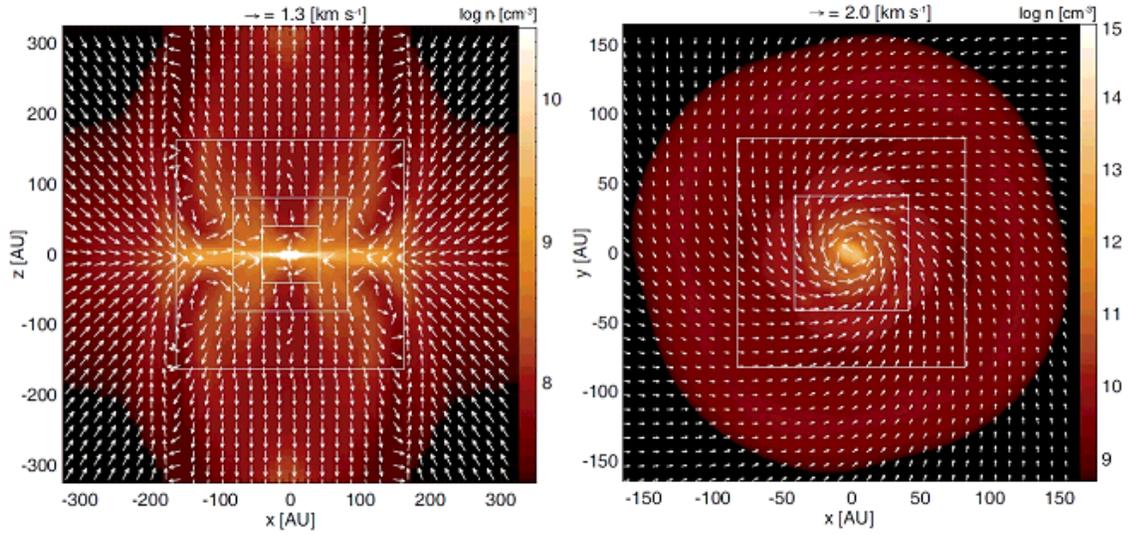}
\caption{
Density distribution (color) on $y=0$ plane (left panel) and $z=0$ plane (right panel) for model A05 at $\tc=3781$\,yr after protostar formation.
Arrows indicate the velocity.
The left panel has a box size of $320$\,AU ($l=10$, where $l$ is the grid level), while the right panel has a box size of $160$\,AU ($l=11$).
}
\label{fig:6}
\end{figure}

Unlike model A03, model A05 shows no fragmentation in the circumstellar disk by the end of the calculation.
Figure~\ref{fig:6} shows the density distribution on the $y=0$ (left) and $z=0$ (right) planes  $\tc=3781$\,yr after protostar formation for model A05.
As in model A03, model A05 also shows the outflow driven by the circumstellar disk.
The outflow appears $\sim700$\,yr before protostar formation and reaches $\sim4400$\,AU at the end of the calculation.
In the right panel, the ring-like density gap is seen in the range of $40\,{\rm AU}\, \lesssim r \lesssim 80\,$AU.
In this range, the gas is largely blown away from the circumstellar disk by the outflow.

In the right panel, a non-axisymmetric structure is seen in the proximity of the protostar ($r\lesssim40$\,AU), while nearly axisymmetric structures are seen in the outer disk region ($r\gtrsim40$\,AU).
The magnetic field is well-coupled with neutrals in the outer disk region because the gas density is lower than $n\lesssim 10^{11}\cm$ where Ohmic dissipation is not effective.
Thus, the angular momentum is effectively transferred by magnetic braking in the outer disk region, and the gas in such a region can fall into the inner disk region.
On the other hand,  the magnetic field is significantly dissipated by the Ohmic dissipation in the inner disk region ($r\lesssim40$\,AU), because the density in such a region exceeds $n\gtrsim10^{11}\cm$.
Thus, in the inner disk region, the angular momentum transfer by the magnetic field is not effective, and the excess angular momentum and accumulated mass cause the gravitational instability to induce the non-axisymmetric structure \citep{walch09a,machida10}.
However, since the increase rate of the disk mass for model A03 is considerably smaller than that for model A05 (see, \S\ref{sec:mass1}), the circumstellar disk for model A05 is not sufficiently massive to induce fragmentation (see, \S\ref{sec:property}).
For model A05, fragmentation and formation of substellar-mass object do not occur until the end of the calculation.
As a result, since the circumstellar disk is not disturbed by the fragments, the protostellar and circumstellar-disk mass moderately increase, unlike in the case of model A03 (see, \S\ref{sec:mass1}).

\subsection{Masses of Protostar and Circumstellar Disk}
\label{sec:mass1}
\begin{figure}
\includegraphics[width=150mm]{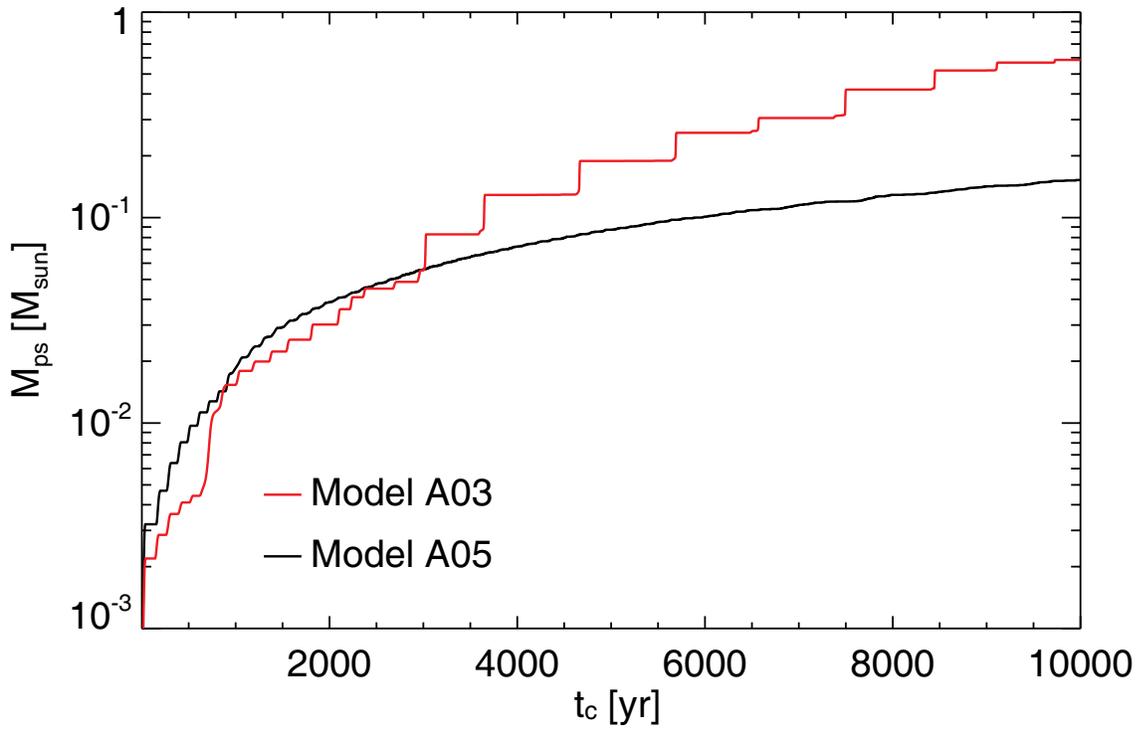}
\caption{
Protostellar mass for models A03 (red) and A05 (black) against time elapsed after protostar formation.
}
\label{fig:7}
\end{figure}

Figure~\ref{fig:7} shows the evolution of protostellar mass against time after protostar formation.
The protostellar mass $M_{\rm ps}$ is derived as the mass falling into the sink.
At its formation, the protostar has a mass of $M_{\rm ps} \sim 10^{-3}\msun$, as expected in spherically symmetric calculations \citep{larson69,masunaga00}.
In the main accretion phase, protostars increases their mass by gas accretion and reach $M_{\rm ps} = 0.59\msun$ (model A03) and  $0.15\msun$ (model A05), respectively, at the end of the calculation.
In Figure~\ref{fig:7}, the step-like increase of the protostellar mass for model A03 is caused by the falling of the planet.
Since the planet and its envelope have a total mass of $\sim6-27\mjup$ (see, Table~2), the protostar drastically increases its mass at the moment of the falling.
In both models, the protostars have almost the same mass for $\tc \lesssim 3000$\,yr, while the protostar for model A03 is  more massive than that for model A05 for $\tc \gtrsim 3000$\,yr.
The difference in  protostellar mass is caused by the difference in the rate of mass accretion onto the protostar.
The different mass accretion rates are owing to different efficiencies of the angular momentum transfer in the circumstellar disk. 
The higher rate of mass accretion onto the protostar is realized in a more massive circumstellar disk, because a massive disk tends  to show gravitational instability that effectively transfers the angular momentum outward, and increases the rate of mass accretion onto the protostar.

\begin{figure}
\includegraphics[width=150mm]{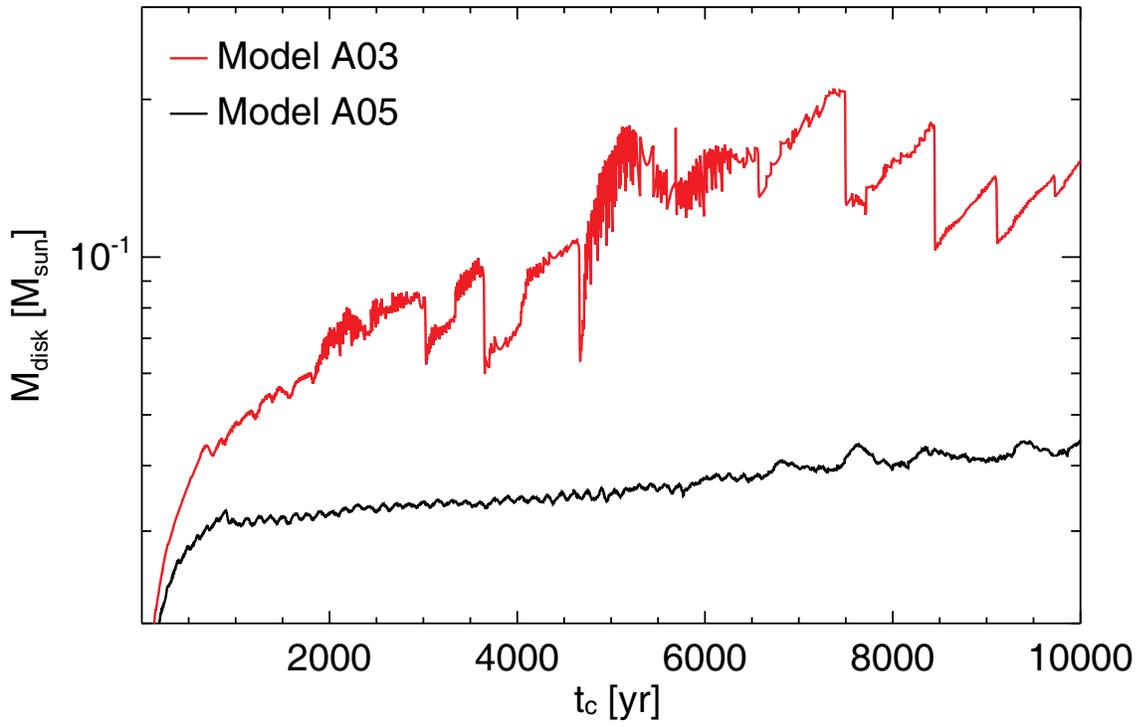}
\caption{
Mass of circumstellar disk for models A03 (red) and A05 (black) against elapsed time after protostar formation.
}
\label{fig:8}
\end{figure}

Figure~\ref{fig:8} shows the time evolution of the circumstellar-disk mass that is estimated in the same manner as in \citet{machida10}.
The figure indicates that the disk mass for model A03 is significantly greater than that for model A05 at the epoch just after protostar formation. 
The circumstellar disk for model A05 gradually increases its mass until the end of the calculation and reaches $M_{\rm disk}\sim 0.042\msun$ for $\sim10^4$\,yr.
The average increase rate of disk mass for model A05 is $\dot{M}_{\rm disk}\sim4.2\times10^{-6}\mdot$.
On the other hand, the circumstellar disk mass for model A03 reaches $M_{\rm disk} \sim0.15\msun$ for $\sim10^4$\,yr with an average increase rate of $\dot{M}_{\rm disk}\sim1.5\times10^{-5}\mdot$.
Thus, considering that the initial stability of the cloud core is closely related to the mass accretion rate (see, \S\ref{sec:mass}), the increase rate of the disk mass for model A03 is about one order of magnitude higher than that for model A05.
Note that it is difficult to accurately estimate the total mass of accreting gas onto the circumstellar disk because a fraction of the circumstellar-disk mass falls onto the protostar with time.
For model A03, the disk mass shows a high time variability that is caused by the time variable mass accretion onto the protostar (see, \S\ref{sec:mdot}); the spike-like decrease of the disk mass for model A03 corresponds to the epoch at which the planet falls onto the protostar.

\subsection{Rate of Mass Accretion onto Protostar}
\label{sec:mdot}
\begin{figure}
\includegraphics[width=150mm]{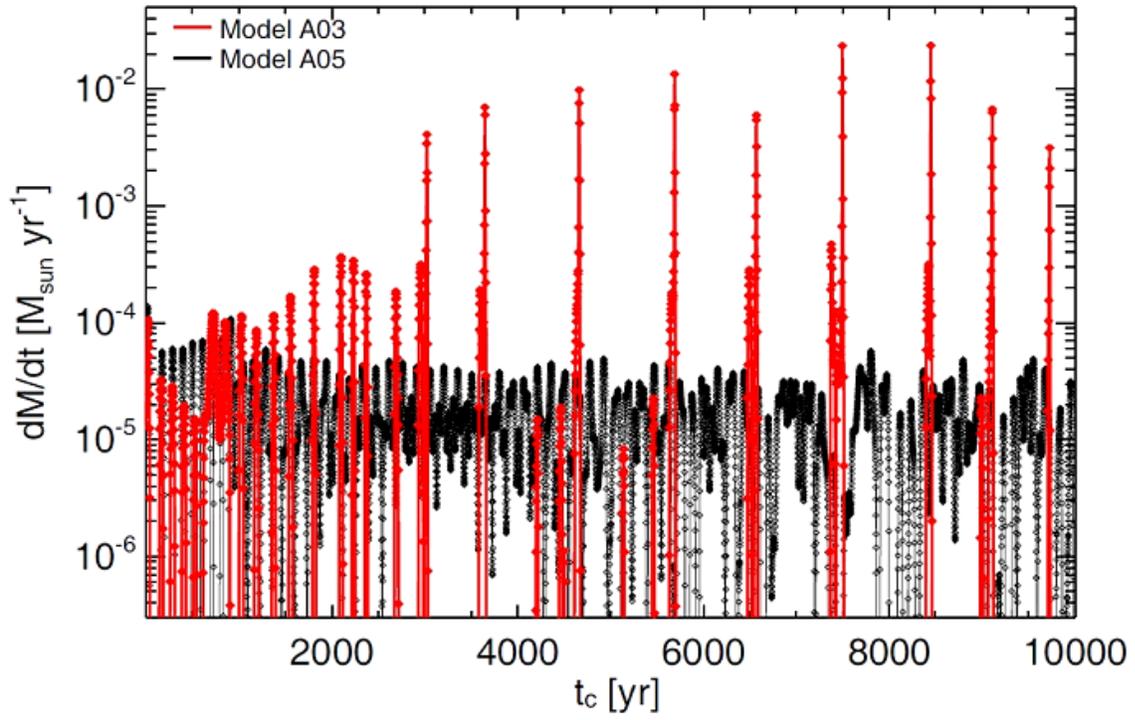}
\caption{
Mass accretion rate for models A05 (black) and A03 (red) against time elapsed after protostar formation.
}
\label{fig:9}
\end{figure}

Since the gas accretes onto the protostar mainly through the circumstellar disk in the main accretion phase, the mass accretion rate depends on disk properties such as disk mass, spiral structure, and existence of fragments.
Figure~\ref{fig:9} shows the mass accretion rates for $\sim10^4$\,yr after protostar formation.
We estimated the mass falling into the sink every one year and defined it as the rate of mass accretion onto the protostar, $\dot{M}_{\rm ps}$.
Thus, the time resolution of the accretion rate in Figure~\ref{fig:9} is one year, while the time-step of calculations is $\Delta t \simeq 0.001-0.01$\,yr.
Therefore, we integrated the falling mass every $\sim100-1000$ time-steps.

In Figure~\ref{fig:9}, the mass accretion rate for model A05 gradually decreases, oscillating between $\dot{M}\simeq10^{-4}$ and $\dot{M} \sim 10^{-6}\mdot$, while model A03 shows intermittent gas accretion with multiple peaks.
For model A05, it is considered that the spiral structure in the inner disk region induces the non-steady accretion flow.
The spiral structure can transfer the angular momentum outward and move the gas inward.
The range of the accretion rate in this model almost corresponds to the theoretical prediction that is expressed as
\begin{equation}
\dot{M}=f_{\rm m}\, \dfrac{c_s^3}{G} \simeq 1.9\times10^{-6}\, f_{\rm m} \, \mdot,
\label{eq:mdot}
\end{equation}
where $f_{\rm m}$ is a numerical factor (e.g., $f_{\rm m}=0.975$ for \citealt{shu77}, $f_{\rm m}= 46.9$ for \citealt{hunter77}), and $c_s = 200$\,m\,s$^{-1}$ is adopted.
The gas falls onto the circumstellar disk first, and the rate of gas accretion onto the circumstellar disk is $\sim4\times10^{-6}\mdot$ on average, as shown in Figure~\ref{fig:8}.
In summary, the rate of mass accretion onto the protostar almost corresponds to the mass accretion onto the circumstellar disk for model A05.

On the other hand, model A03 shows intermittent mass accretion caused by the large-scale perturbation of the disk and falling of planets.
As shown in Figure~\ref{fig:4}, the first-generation planets form at $r\sim 6.8$\,AU and orbit in the range of $2$ to $40$\,AU until they fall onto the planet.
They disturb the circumstellar disk and transfer the angular momentum outward.
Thus, the mass accretion rate is linked to the orbital motion of the planet especially for $\tc \lesssim 3000$\,yr, i.e., before the first-generation planets fall onto the protostar.
After the first-generation planets disappear, planets of subsequent generations  (2nd to 9th) appear at $r\simeq20-50$\,AU from the protostar. 
They orbit in the range of from 5 to 50\,AU and approach the protostar within $r<5$\,AU only before falling onto the protostar.
Thus, just before falling, these planets can disturb the inner disk region and promote mass accretion onto the protostar.
The peak mass accretion rate for $\tc\gtrsim3000$\,yr corresponds to the epoch in which a planet approaches and falls onto the protostar.
In addition, since not only the gas in the circumstellar disk but also the planet itself falls onto the protostar, the mass accretion rate at its peak for a model with planets is much larger than the theoretical prediction.
In Figure~\ref{fig:9}, there are eight peaks of the mass accretion rate for $\tc \gtrsim 3000$\,yr, corresponding to epochs of falling of planets.
In such an epoch, the protostar has a very high mass accretion rate of $\dot{M}_{\rm ps} \gtrsim 10^{-3}-10^{-2}\mdot$, because a planet with mass of $\sim 10^{-3}-10^{-2}\msun$ falls onto the protostar in a short duration.
Note that this high accretion rate may be smoothed over time if calculated with higher spatial resolution, while it is considered that the total mass falling onto the protostar does not change.

\subsection{Disk Properties for Models with and without Fragmentation}
\label{sec:property}
\begin{figure}
\includegraphics[width=150mm]{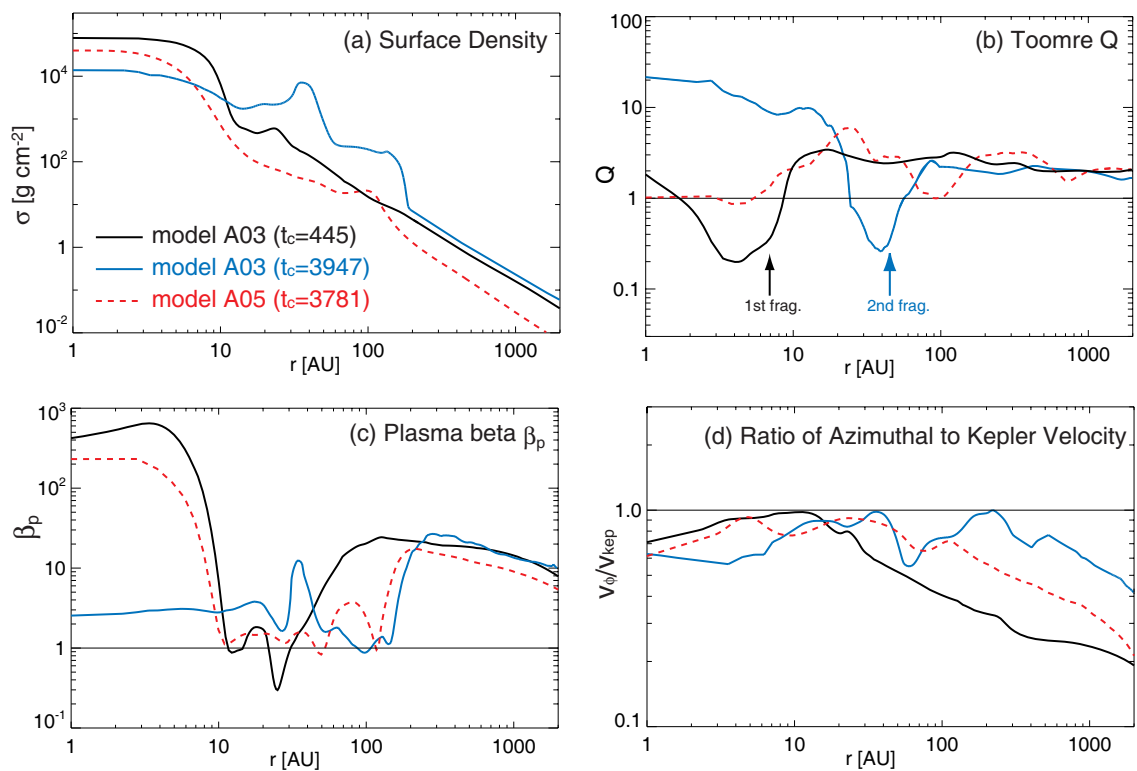}
\caption{
Radial distribution of the surface density ({\it a}), Toomre Q parameter ({\it b}), plasma beta $\beta_{\rm p}$ ({\it c}), and ratio of the azimuthal to the Kepler velocity ({\it d}) for models with A03 ($\tc=445$ and 3917\,yr) and A05 ($\tc=3781$\,yr).
}
\label{fig:10}
\end{figure}

Figure~\ref{fig:10} shows the radial distribution of the surface density $\sigma$ (Fig.~\ref{fig:10}{\it a}), Toomre $Q$ parameter (Fig.~\ref{fig:10}{\it b}, \citealt{toomre64}), plasma $\beta_p$ (Fig.~\ref{fig:10}{\it c}), and ratio of the azimuthal to Kepler velocity $v_\phi/v_{\rm kep}$ (Fig.~\ref{fig:10}{\it c}).
These quantities are azimuthally averaged.
The azimuthal-averaged surface density $\sigma$ [$\equiv \sigma (r)$] is derived as
\begin{equation}
\sigma (r) = \dfrac{dr \int_0^{2\pi } \sigma_{\rm s} (r, \theta)\, r_{\rm c}\,  d\theta}{dS},\\
\end{equation}
where,
\begin{equation}
\sigma_{\rm s} (r, \theta) = \int^{z < z_{\rm cri}} \rho(r, \theta, z)\, dz,\\
\label{eq:sigmas}
\end{equation}
and
\begin{equation}
dS = dr \int_0^{2\pi} r_{\rm c}\, d\theta,
\end{equation}
where $r_c$ is the cylindrical radius, and $dr=ds(l)$ (the cell width of $l$th-level grid) is adopted.
To determine $z_{\rm cri}$ in equation~(\ref{eq:sigmas}), we estimated the disk height by eyes and $z_{\rm cri}$=10\,AU is adopted.
We confirmed that the surface density $\sigma$ hardly depend on $z_{\rm cri}$ in the range of $z_{\rm cri}>5$\,AU.
In Figures~\ref{fig:10}{\it b} and {\it c},  Q parameter and plasma $\beta_{\rm p}$ are integrated along $z-$axis and azimuthally averaged.
To estimate Q parameter, firstly, we integrated the mass-weighted sound speed $c_s$ and epicyclic frequency $\kappa$ along $z-$axis as
\begin{equation}
c_{\rm s,s} (r, \theta) = \dfrac{\int^{z < z_{\rm cri}} c_s(r, \theta, z)\, \rho(r, \theta, z)\,dz}{\sigma_{\rm s}(r, \theta)},
\label{eq:csr}
\end{equation}
and
\begin{equation}
\kappa_{\rm s} (r, \theta) = \dfrac{\int^{z < z_{\rm cri}} \kappa (r, \theta, z)\, \rho(r, \theta, z)\,dz}{\sigma_{\rm s}(r, \theta)}.
\label{eq:kappar}
\end{equation} 
Then, using equations~(\ref{eq:sigmas}), (\ref{eq:csr}) and (\ref{eq:kappar}), we derived the azimuthal-averaged Q parameter as
\begin{equation}
Q (r) = \dfrac{dr \int_0^{2 \pi} Q(r, \theta)\, r_c\, d\theta}{dS},
\end{equation}
where
\begin{equation}
Q(r, \theta) = \dfrac{c_{\rm s,s}(r, \theta)\, \kappa_s (r, \theta)}{\pi G\, \sigma_r (r, \theta)}.
\end{equation}
The azimuthal-averaged plasma $\beta_{\rm p} (\equiv 8\pi c_s^2 \rho/B^2$) is also derived in a similar way to Q parameter as
\begin{equation}
\beta_{\rm p} (r) = \dfrac{dr \int_0^{2 \pi} \beta_{\rm p, s} (r, \theta)\, r_{\rm c}\, d\theta}{dS},
\end{equation}
where
\begin{equation}
\beta_{\rm p,s}(r, \theta) = \dfrac{\int^{z < z_{\rm cri}} \beta_{\rm p} (r, \theta, z)\, \rho(r, \theta, z)\,dz}{\sigma_{\rm s}(r, \theta)}.
\end{equation}
In addition, the ratio of azimuthal-averaged $v_\phi$ to the Kepler velocity on the equatorial plane is plotted in Figure~\ref{fig:10}{\it d}.
The black lines in Figure~\ref{fig:10} are the radial quantities at the epoch just before the first fragmentation ($\tc=445$\,yr) for model A03, while blue lines are those at the epoch just before the second fragmentation after the first fragments falling onto the protostar ($\tc=3947$\,yr) for model A03.
The red-dotted lines are quantities at almost the same epoch ($\tc=3781$\,yr) as blue lines but for model A05.

The black lines in Figure~\ref{fig:10}{\it a} and {\it b} show that, for $r<10$\,AU, the surface density has a maximum and $Q$ parameter becomes $Q<1$.
Just after this epoch, fragmentation occurs at $r=6.8$\,AU where  $Q<1$.
The black lines in Figures~\ref{fig:10}{\it c} and {\it d} show that plasma $\beta_{\rm p}$ suddenly rises at $r \sim 10$\,AU inside which the flow moves azimuthally with near the Keplerian velocity.
Since the gas (surface) density is sufficiently high in the region of $r<10$\,AU, the Ohmic dissipation becomes effective and magnetic field considerably dissipates.
In general, fragmentation hardly occurs when the neutral gas is strongly coupled with the magnetic field (or ions), because the magnetic field suppresses fragmentation \citep{hennebelle08b,machida08b}. 
Instead, when $Q<1$, fragmentation is easy to occur in the magnetically inactive region where the Ohmic dissipation weakens the magnetic field.
In summary, at this epoch, fragmentation can occur in the region of $r<10$\,AU, because $Q<1$ is realized owing to the higher surface density and very weak magnetic field (or higher $\beta_{\rm p}$) does not suppress fragmentation.

In the region of $r<10$\,AU, the surface density at $\tc=3947$\,yr (blue line) is lower than that at $\tc=445$\,yr (black line).
This is because, in the inner disk region,  the angular momentum  is transferred by the orbital motion of the first-generation  planets and the gas falls onto the protostar.
In addition, planets contribute to the gap formation around the central protostar as seen in \citet{vorobyov10}.
Thus, at $\tc=3947$\,yr, the surface density near the protostar lowers and $Q$ parameter exceeds $Q>1$ for $r\lesssim30$\,AU (blue line of Fig.~\ref{fig:10}{\it b}).
Since the lower (surface) density makes the Ohmic dissipation ineffective, lower plasma $\beta_{\rm p}$ ($\beta_{\rm p} \sim2-4$) is realized in this region ($r\lesssim 40$\,AU).
Therefore, the inner disk region becomes magnetically active and the angular momentum is  transferred also by the magnetic braking.
Figure~\ref{fig:10}{\it d} shows that azimuthal velocity to the Kepler velocity at $\tc=3947$\,yr is lower than that at $\tc=445$\,yr.
The lower surface density and higher value of plasma $\beta_{\rm p}$ are not adequate to induce fragmentation.
On the other hand, the surface density and plasma $\beta_{\rm p}$ at $\tc=3947$\,yr have a local peak at $r\sim30-40$\,AU (blue lines of Figs.~\ref{fig:10}{\it a} and {\it c}).
This is because, in this region, the radially inward motion of the flow is suppressed by the first-generation planet and the gas is accumulated.
In such high-density gas region, the Ohmic dissipation becomes effective owing to the higher surface density.
Thus, the magnetic braking becomes ineffective and  azimuthal velocity approaches the Kepler speed (blue line of Fig.~\ref{fig:10}{\it d}).
Therefore, the centrifugal barrier further amplifies the surface density.
Finally, $Q$ parameter becomes $Q<1$ in the range of $20\,{\rm AU} \lesssim r \lesssim 60\,{\rm AU}$ (Fig.~\ref{fig:10}{\it b}),  and fragmentation occurs to form the second-generation planets at $r\simeq46$\,AU as seen in Figure~\ref{fig:1}{\it g}.

In Figures~\ref{fig:10}{\it a} and {\it b}, model A05 (red-dotted line) has a relatively lower surface density because this model has a lower mass accretion rate.
Therefore,  $Q$ parameter never reach $Q\ll1$, and fragmentation does not occur until the end of the calculation for model A05.
In addition, the lower surface density increases the magnetically active area (Fig.~\ref{fig:10}{\it c}) and stronger outflow appears as seen in Figure~\ref{fig:6}.
Thus, the excess angular momentum is mainly transferred by the protostellar outflow and magnetic braking for model A05, while the excess angular momentum contributes to fragmentation for model A03.

In summary, the initial stability (or mass accretion rate) is related not only the surface density (or mass of the circumstellar disk) but also the magnetic activity in the circumstellar disk.
The higher accretion rate realizes the higher surface density (i.e., lower $Q$) and has a large magnetically inactive area where the angular momentum is not transferred by the magnetic effects.
Both effects promote fragmentation and planet formation in the circumstellar disk.
Instead, since the lower accretion rate has a lower surface density and large magnetically active area, fragmentation and planet formation tend to be suppressed.

\subsection{Protostellar Outflow Driven by Circumstellar Disk}
\label{sec:outflow}
\begin{figure}
\includegraphics[width=150mm]{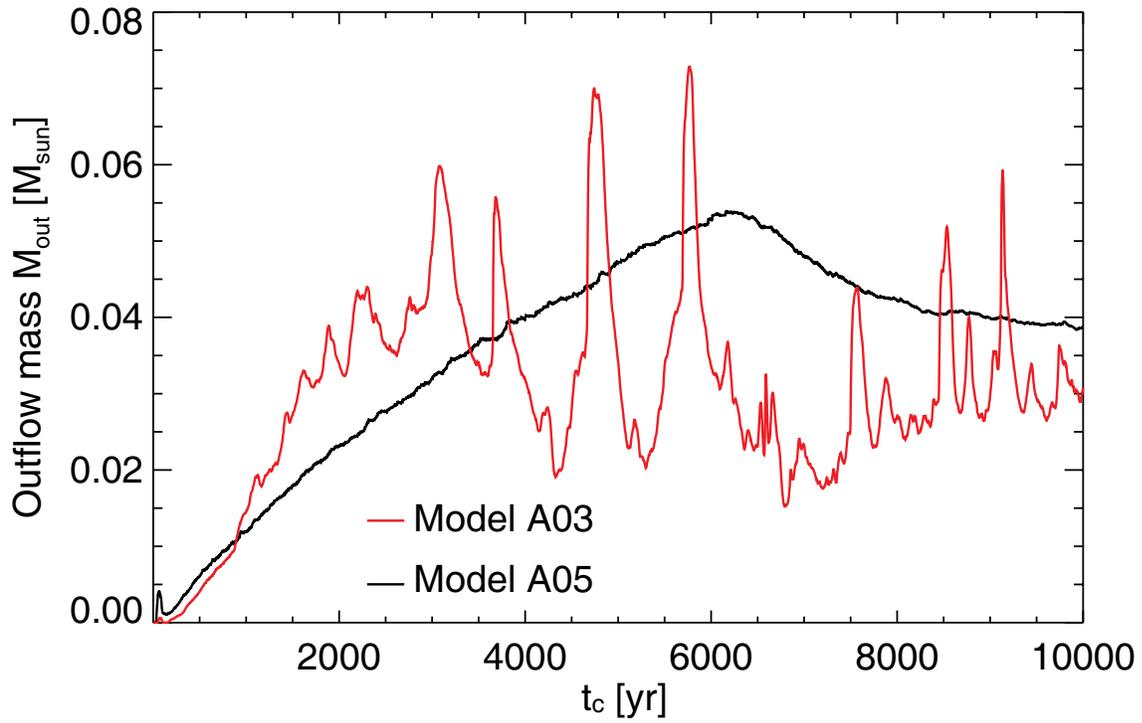}
\caption{
Outflowing mass for models A05 (black) and A03 (red) against time elapsed after  protostar formation.
Outflow is defined as gas with radial velocity exceeding sound speed $v_r > c_{\rm s,0}$ inside $r<\rcri$.
}
\label{fig:11}
\end{figure}

The condition of the circumstellar disk also affects  protostellar outflow.
Figure~\ref{fig:11} shows the mass outflowing from the protostellar system (i.e., protostar + circumstellar disk system) for both models.
To measure the outflowing mass, we integrated the gas having $v_r>c_{s,0}$ inside the initial radius of cloud core ($r<\rcri$), where $v_r$ is the radial velocity.
The outflowing mass for model A05 gradually increases for $\tc\lesssim 6000$\,yr and decreases for $\tc\gtrsim6000$\,yr.
Since the freefall timescale of the initial cloud core is $\sim1.8\times10^4$\,yr, it is expected that the outflow weakens after passing the peak and disappears on a freefall timescale.
In addition, the outflow extends far beyond the initial radius of cloud core for $\tc\gtrsim5000$\,yr.
Thus, we underestimated the outflowing mass for $\tc\gtrsim5000$\,yr because we integrated the outflowing mass only inside $r<\rcri$.
To conclusively estimate the outflowing mass and total ejected mass from the initial cloud core, more long-term calculations are necessary.
However, since such calculations are beyond the scope of this study, we do not comment on them any more.
At its peak, the outflowing mass is about $0.05\msun$ which is comparable to the protostellar mass $M_{\rm ps}\simeq 0.1\msun$ (see Fig.~\ref{fig:7}) at the same epoch.
Thus, a large fraction of the infalling gas is blown away from the circumstellar disk by the protostellar outflow for model A05.

Unlike model A05, in Figure~\ref{fig:11}, the outflow mass for A03 shows high time variability, indicating that outflow is unsteadily driven by the circumstellar disk.
In summary, Figure~\ref{fig:11} indicates that steady outflow is driven by the less massive circumstellar disk without fragments (model A05), whereas intermittent outflow is driven by the more massive circumstellar disk with fragments (model A03).
Note that Figure~\ref{fig:11} shows only the outflowing mass having the high-velocity component of $v>c_{s,0}$, and ignores the low-velocity component of the outflow.

\begin{figure}
\includegraphics[width=150mm]{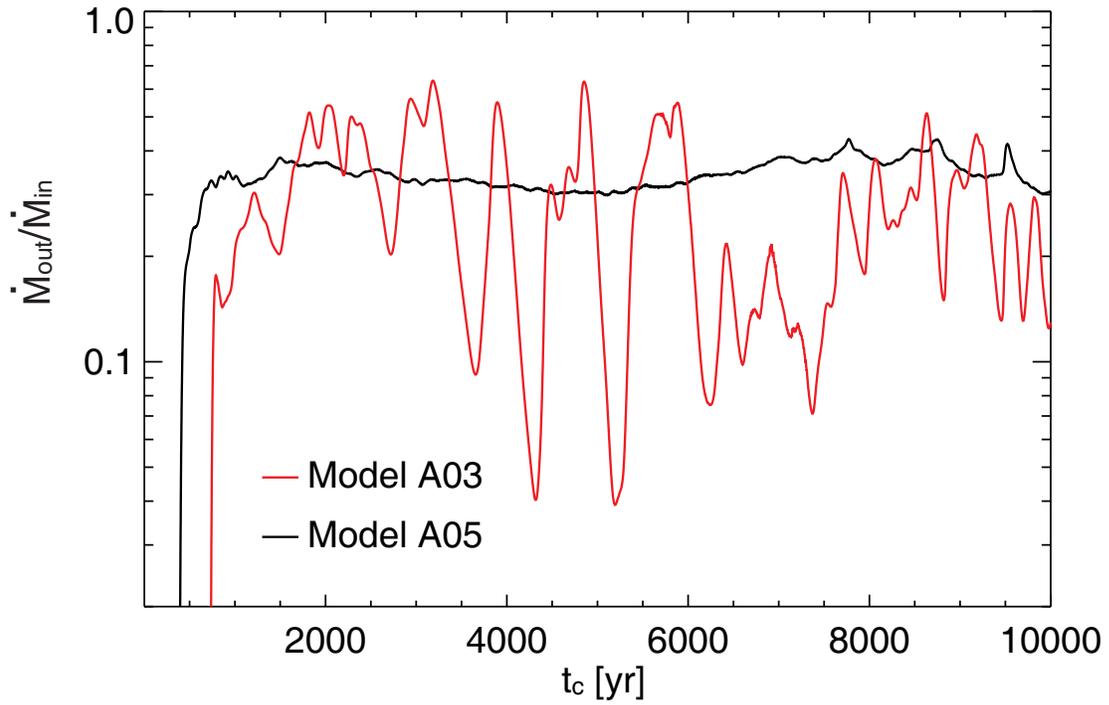}
\caption{
Ratio of outflowing to inflowing mass for models A05 (black) and A03 (red) against time elapsed  after protostar formation.
The outflowing mass is estimated as the mass ejected from the $l=9$ grid, and the inflowing mass is estimated as mass entering the  $l=9$ grid.
}
\label{fig:12}
\end{figure}

Figure~\ref{fig:12} shows the ratio of the outflowing to infalling mass.
To investigate the outflow efficiency, we calculated the outflowing/infalling mass rate on a $l=9$ grid surface, which covers the whole circumstellar disk with a grid size of $\sim600$\,AU.
The box size of $l=9$ grid is $\sim700$\,AU.
According to \citet{tomisaka02}, we estimated the outflow $\dot{M}_{\rm out}$ and inflow $\dot{M}_{\rm in}$ masses as 
\begin{equation}
\dot{M}_{\rm out} = \int_{\rm boundary\, of \, l=9} \rho\, {\rm max}[\vect{v} \cdot \vect{n},0]\, dS,
\end{equation}
and
\begin{equation}
\dot{M}_{\rm in} = \int_{\rm boundary\, of \, l=9} \rho\, {\rm max}[\vect{v} \cdot \vect{-n},0]\, dS,
\end{equation}
where $\vect{n}$ is the unit vector outwardly normal to the surface of the $l=9$ grid.
In Figure~\ref{fig:12}, model A05 keeps an almost constant ratio, $\dot{M}_{\rm out}/\dot{M}_{\rm in}\sim0.3$, until the end of the calculation.
Thus, about 30\% of the infalling mass is ejected by the outflow.
In this model, the outflow is driven by the outer disk region where the axisymmetric structure is kept, while no outflow appears in the inner disk region where a considerably weak magnetic field is realized by Ohmic dissipation and a non-axisymmetric structure appears, as seen in Figure~\ref{fig:6}.
In the outer disk region, the disk mass is blown away by the outflow, and effective magnetic braking moves the gas inward.
Therefore, such a region does not maintain mass sufficient to induce the gravitational instability.
As a result, the steady disk region drives the steady flow for model A05.

On the other hand, the outer disk region for model A03 does not remain stable, because the increase rate of the disk mass (or the mass accretion rate onto the disk) for model A03 is much larger than that for model A05.
Although the magnetic braking and mass ejection by the outflow are effective in the outer disk region for model A03 also, they cannot transfer sufficient mass  to suppress the gravitational instability and subsequent fragmentation owing to the high mass accretion onto the disk.
Thus, multiple planets appear in the disk.
Oscillation of the infall-outflow mass ratio for model A03 is shown in Figure~\ref{fig:12}.
Comparison of Figure~\ref{fig:4} with Figure~\ref{fig:12} indicates that this oscillation is related to the orbital motion of the planet.
It is considered that the magnetic field lines and surface density of the circumstellar disk are disturbed by the planet's orbital motion, which causes highly time-variable outflow, as shown in Figure~\ref{fig:12}.
The time-variable protostellar outflow may be a clue for detecting a planet in a very early phase of star formation.

\section{Discussion}
\subsection{Can We Detect Planets or Substellar-Mass Objects in the Main Accretion Phase?}
\begin{figure}
\includegraphics[width=150mm]{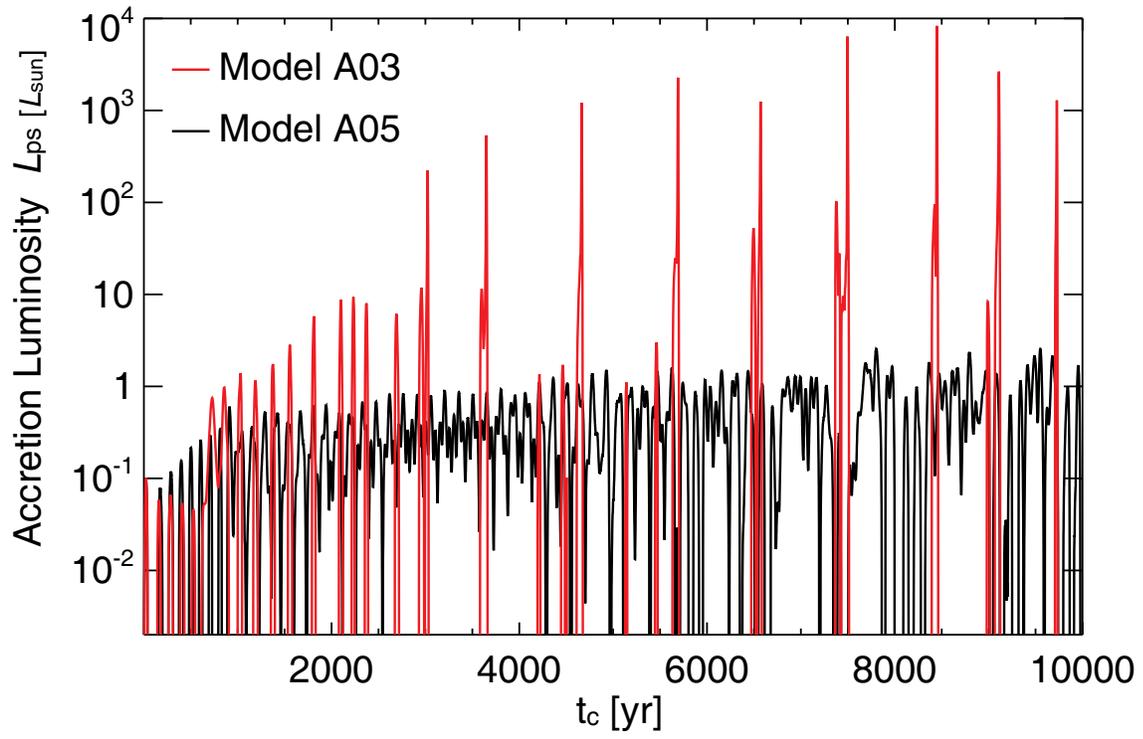}
\caption{
Accretion luminosity of protostar for models A05 (black) and A03 (red) against time elapsed time after protostar formation.
}
\label{fig:13}
\end{figure}

So far, it is considered that planets form in relatively quiet circumstellar disks after the gas accretion onto the circumstellar disk has almost ended.
In this study, however, we showed the possibility of the formation of planets or substellar-mass objects by gravitational instability in a vigorous circumstellar disk during the main accretion phase, in which the circumstellar disk continues to increase its mass by gas accretion from the infalling envelope.
When planets or substellar-mass objects appear in the circumstellar disk, they can affect the rate of accretion onto the protostar.
As described in \S\ref{sec:mdot}, we may verify their existence by observing the time-variability of the rate of accretion onto the central protostar.
As seen in Figure~\ref{fig:9}, the protostar shows a relatively low time-variability of the accretion rate when neither planets nor substellar-mass objects exist in the circumstellar disk, while it shows a high time-variability of the accretion rate when fragmentation occurs and planets or substellar-mass objects appear. 
Figure~\ref{fig:13} shows the time variation of accretion luminosity of the protostar, which is defined as
\begin{equation}
L_{\rm ps} = G\, \dfrac{M_{\rm ps}\, \dot{M}_{\rm ps}}{r_{\rm ps}},
\end{equation}
where we adopted the protostellar mass $M_{\rm ps}$ and mass accretion rate $\dot{M}_{\rm ps}$ as values derived from calculations at each time and used a constant protostellar radius $r_{\rm ps} = 2\rsun$ for simplicity.
The figure indicates that the protostar for model A05 has an accretion luminosity in the range of $0.01 \lesssim  L_{\rm ps}/\lsun \lesssim 1$ for $\sim10^4$\,yr after protostar formation.
The accretion luminosity for this model gradually increases for a longer timescale of $\sim10^4$\,yr, while it oscillates with a period of $\sim100$\,yr, which  corresponds to the Kepler timescale in the region of a developing spiral structure.
The Kepler timescale $\tau_{\rm k}$ can be expressed as 
\begin{equation}
\tau_{\rm k} = 2\pi \left( {\dfrac{r^3}{G M_{\rm p}}}  \right)^{1/2} \simeq 100 \left( \dfrac{r}{10\,{\rm AU}} \right)^{3/2} 
\left( \dfrac{0.1\msun}{M_{\rm ps}}  \right)^{-1/2}\, {\rm yr}.
\end{equation}
As shown in Figure~\ref{fig:6}, the non-axisymmetric structure appears at $r\lesssim 10-40$\,AU, which corresponds to the magnetic non-active zone \citep{machida07}.
Thus, it is natural that the protostar shows time-variable accretion luminosity with a period of $\sim100$\,yr, because the angular momentum transfer that is closely related to the rate of mass accretion onto the protostar is caused  by the spiral structure in the circumstellar disk.
The variation in amplitude of the accretion luminosity for a short duration ($\sim100$\,yr) is about one order of magnitude.

On the other hand, the accretion luminosity for model A03 is qualitatively and quantitatively different from that for model A05.
The protostar for model A03 repeatedly shows a strong brightening.
In this model, the accretion luminosity suddenly increases to reach $\sim10-10^4\lsun$ when planets or substellar-mass objects approach the protostar or fall onto the protostar, while the luminosity is lower, $\lesssim 0.1-0.01\lsun$ when planets or substellar-mass objects orbit far from the protostar.
Thus, the strong brightness may be a clue for detecting planets or sub-stellar-mass objects.

\begin{figure}
\includegraphics[width=150mm]{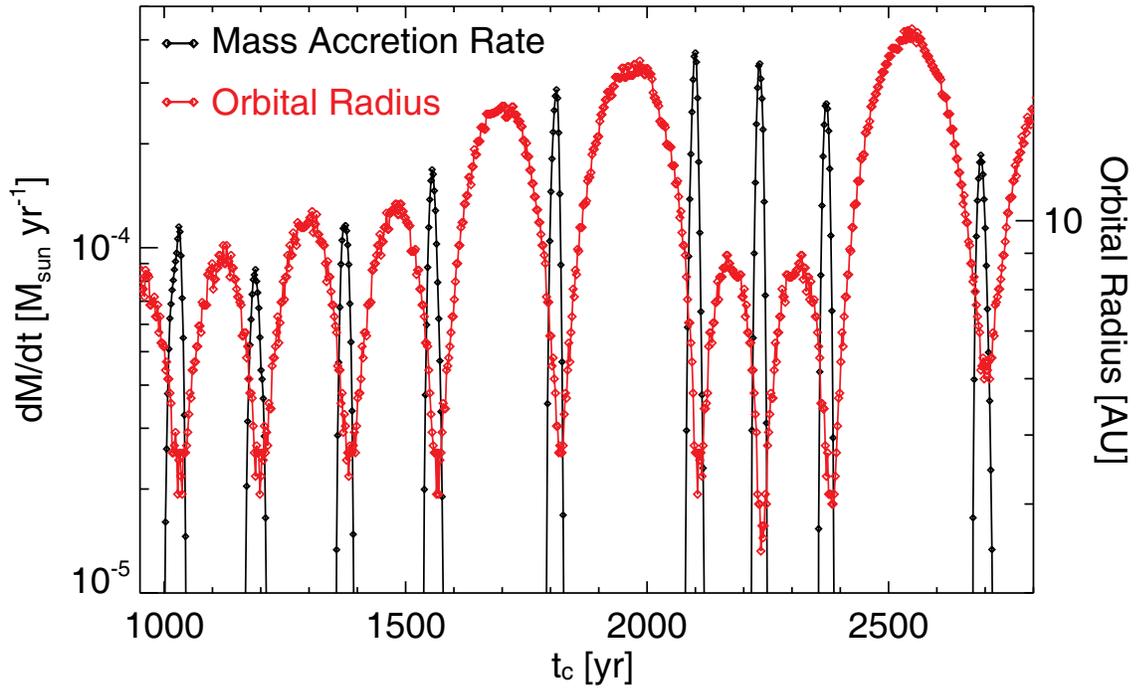}
\caption{
Mass accretion rate (left axis; black line and dots) and orbital radius of the most massive fragment (right axis; red line and dots) for model A03 during $\tc=1000 - 2800$\,yr after protostar formation.
The perihelion of the planet corresponds to the peak of the mass accretion rate.
}
\label{fig:14}
\end{figure}

In addition, when the planets or substellar-mass objects orbit in the inner disk region of $\lesssim 10-20$\,AU, their orbital motion is linked to the accretion rate (or accretion luminosity).
The accretion rate and orbital radius for  $\tc=1000-2800$\,yr for model A03 are plotted in Figure~\ref{fig:14}.
In this time period, the first-generation planet orbits around the protostar for $\sim10$ times.
The figure indicates that the orbital motion of planets or substellar-mass objects clearly synchronizes with the mass accretion rate.
The orbiting objects disturb the circumstellar disk and promote mass accretion onto the protostar.
Since the circumstellar disk has a higher density and shorter timescale in the proximity of the protostar, a high mass accretion rate is realized when the planets or substellar-mass objects approach the protostar.
As a result, the mass accretion rate is closely related to the orbital motion of planets and substellar-mass objects.
We may estimate and verify the orbital motion of planets in the circumstellar disk by observing the time-variable mass accretion.
Note that we ignored the effect of the infalling envelope when the accretion luminosity is estimated in Figure~\ref{fig:13}.
Note also that, to quantitatively estimate the accretion luminosity, we have to consider the infalling envelope that weakens the luminosity from the protostar.

\subsection{Mass Accretion Rate onto Disk and Masses of Fragments}
\label{sec:mass}
In this study, we showed that fragmentation and subsequent formation of planet and sub-stellar mass object can occur in a massive circumstellar disk, while no fragmentation occurs in a relatively less massive disk.
The mass of the circumstellar disk or the mass increase rate of the circumstellar disk is related to the initial stability of cloud core.
We parameterized the ratio of thermal energy to gravitational energy, $\alpha_0$, which is related to the mass accretion rate.
In general, the mass accretion rate is described as
\begin{equation}
\dot{M} = \dfrac{M_{\rm J}}{t_{\rm ff}} \propto \dfrac{c_s^3}{G},
\end{equation}
where $M_{\rm J}$ is the Jeans mass and $t_{\rm ff}$ is the freefall timescale.
In this study, we made an equilibrium sphere (i.e., critical BE sphere) and increased the density (or mass)  by a factor of $f$ ($=0.84/\alpha_0$, see eq.~[\ref{eq:alpha}]) to induce the collapse of the cloud core.
Thus, the mass accretion rate in our setting can be  described as
\begin{equation}
\dot{M} = \dfrac{M_{\rm cl}}{t_{\rm cl,ff}} = \alpha_0^{-3/2}\, \dfrac{c_s^3}{G},
\end{equation}
where the initial mass of cloud core $M_{\rm cl}$ is proportional to the density enhancement factor $f$ and Jeans mass $M_J$ as 
$
M_{\rm cl}\propto f\, M_{\rm J} \propto \alpha^{-1}\, M_{\rm J},
$
and the freefall timescale of the cloud core $t_{\rm ff,cl}$ is proportional to 
$
t_{\rm ff,cl}\propto f^{-1/2}\, t_{\rm ff,0} \propto \alpha^{1/2}\, t_{\rm ff,0},
$
where $t_{\rm ff,0}$ is the freefall timescale of a critical BE sphere.
Thus, a more unstable cloud core that has a smaller $\alpha_0$ has a larger mass accretion rate and tends to show fragmentation due to gravitational instability.
In reality, the mass increase rate for the model A03 is about three times higher  than that for the model A05.
Thus, the fragmentation and subsequent planets or substellar-mass objects may appear in relatively unstable cloud cores.

In general, however, the initial clouds are characterized by three parameters, $\alpha_0$, $\beta_0$, and $\gamma_0$, as described in \S\ref{sec:setting}.
Although we fixed the angular velocity ($\beta_0$)  and magnetic field strength ($\gamma_0$)  of the initial cloud core in this study, fragmentation may occur in more stable cloud cores (i.e., cloud cores with large $\alpha_0$) with different $\beta_0$ and $\gamma_0$.
As shown in  \citet{hennebelle08b}, in the collapsing cloud core, the magnetic field suppresses fragmentation while the rotation promotes it.
Thus, fragmentation tends to occur in the cloud core with larger $\beta_0$ and smaller $\gamma_0$ \citep{machida08a}.
Moreover, even in a cloud core with a very weak magnetic field, fragmentation in the circumstellar disk may be suppressed by the global spiral structure that effectively transports the angular momentum and lowers the surface density (i.e., increases the Toomre $Q$-parameter).
On the other hand, when the circumstellar disk is strongly magnetized, the angular momentum is transferred by magnetic effects and the spiral structure hardly develops.
Then, the mass accumulates in the inner disk region where Ohmic dissipation significantly weakens the magnetic field, and fragmentation can occur because no global spiral structure develops.
Thus, magnetic field and rotation contribute to both promotion and suppression of fragmentation through a complicated process.
In this study, we showed the possibility of fragmentation in the main accretion phase.
However, to quantitatively determine fragmentation and subsequent formation of planets and substellar-mass objects in the circumstellar disk, we should investigate the evolution of the circumstellar disk from the molecular cloud core in a large parameter space.

In the circumstellar disk, fragmentation occurs and planets or substellar-mass objects appear.
In this study, we resolved the fragmentation process with sufficient spatial resolution \citep{truelove97}.
However, more spatial resolution and further developed numerical techniques may be necessary to determine the final mass of fragments.
At its formation, a fragment's mass almost corresponds to the Jeans mass.
In this study, we assumed the barotropic equation of state, in which the minimum Jeans mass is limited to $\sim5\mjup$.
Thus, fragmentation does not occur with a fragment mass of $M \lesssim 5\mjup$.
If the circumstellar disk cools to reaches lower temperatures, more less-massive fragments may appear.
The barotropic approximation makes it possible to calculate very long-term evolution of the circumstellar disk and to investigate the evolution of the circumstellar disk by the end of the main accretion phase from the prestellar core stage, while radiation hydrodynamic calculations are necessary to qualitatively estimate the fragment mass.

In addition, after fragmentation occurs in the circumstellar disk, we need a higher spatial resolution to investigate the acquisition process of the fragments' mass.
Recent works about a gas giant planet formation showed that the Hill radius of the protoplanet should be resolved with sufficient spatial resolution for accurate  estimate the rate of mass accretion onto the protoplanet \citep{kley01,dangelo03,bate03,machida10b}.
Note that in these calculations, the protoplanet mass was fixed, and the evolution of the circumstellar disk was calculated with a high spatial resolution for a shorter duration after a gas-giant formed.
The Hill radius is expressed as
\begin{equation}
r_{\rm H} = \left( \dfrac{M_{\rm p}}{3M_{\rm ps}}  \right)^{1/3} r_{\rm p},
\end{equation}
where $r_{\rm p}$ is the orbital radius of the planet.
When the protoplanet orbiting the solar-mass protostar has a mass of $M_{\rm p}=10^{-3} M_{\rm ps}$ ($10^{-2} M_{\rm ps}$) and is located at $r=1$\,AU (10\,AU), the Hill radius is $r_{\rm H} \simeq 0.07$\,AU ($1.5$\,AU).
When the Hill radius is resolved with $\gtrsim10$ cells, we need a minimum spatial resolution (i.e., the size of each cell) of $\sim0.007$\,AU ($\sim0.15$\,AU). 
However, calculation with such high resolution needs much computing power, and thus we only calculated the evolution of the circumstellar disk for a very short duration.
In reality, the evolution of the circumstellar disk was calculated for only $\sim1000$\,yr at best in previous studies, while we calculated the evolution of the circumstellar disk from prestellar core stage until $\sim 10^4$\,yr after protostar formation.

The final mass of fragments  is important to determine whether they are planets or binary companions.
To understand both formation and evolution of (gas-giant) planets, we need to calculate the evolution of the circumstellar disk for a longer duration with higher spatial resolution.
However, such calculations are impossible even using present supercomputers because a longer temporal calculation conflicts with higher spatial resolution.
Thus, in this study, we covered the whole prestellar cloud core and investigated the evolution of the circumstellar disk with moderately coarser spatial resolution $\Delta  \simeq 0.6$\,AU, and showed that fragmentation occurs in the circumstellar disk and a planet or substellar-mass object forms.
However, since we need a higher spatial resolution to determine the final mass of the fragment, we did not investigate this point in greater depth.
In this study, we showed the possibility of the formation of planets or substellar-mass objects in a magnetized disk for the main accretion phase, and the precise mass of such objects should be investigated in future studies.

\subsection{Ohmic Dissipation and Ambipolar Diffusion}
\label{sec:ambipolar}
As described in \S \ref{sec:property}, the recurrent planet formation and intermittent outflow shown in this paper are closely related to the effect of magnetic field and its dissipation in the circumstellar disk.
Figure~\ref{fig:10} shows that the magnetic field is well-coupled with the neutral gas in the outer disk region ($r_c > 10-100$\,AU) where a relatively high ionization rate is realized owing to a relatively low gas density.
Thus, the angular momentum in such region is effectively transferred by the magnetic braking and protostellar outflow, and the gas can flow into the inner disk region.
On the other hand, the magnetic field is not well-coupled with the neutral gas in the inner disk region ($r_c < 10-100$\,AU) where the gas density is high and a considerably low ionization rate is realized.
Thus, the angular momentum transfer due to the magnetic effects is not so effective, and the gas flowing from the outer disk region accumulates in the inner disk region around the boundary between the magnetically active and inactive region at $r_c \sim 5-50$\,AU (see, Fig.~\ref{fig:10}).
In that region, 
the surface density continues to increase and fragmentation tends to occur.
In summary, magnetic dissipation is essential in inducing recurrent fragmentation (or planet formation) in the circumstellar disk.

In this study, we have considered only Ohmic dissipation as the dissipation mechanism of the magnetic field. 
This is because Ohmic dissipation is expected to be the most effective mechanism for magnetic field dissipation in high density region where most of the magnetic flux leakage occurs. 
However, ambipolar diffusion may also contribute to the magnetic flux leakage in some intermediate density region.
Recent studies have shown that the Ohmic dissipation is more effective than the ambipolar diffusion in the range of $n\gtrsim10^{12}\cm$, while the ambipolar diffusion dominates in the range of $10^{11}\cm\lesssim n\lesssim10^{12}\cm$ 
\citep{tassis07a,tassis07b,tassis07c,kunz10}.
In the range of $n\lesssim 10^{11}\cm$ the magnetic field is well-coupled with the neutral gas in a magnetically super critical cloud \citep{nakano02,kunz10}. 
Thus, it might be important to study how the inclusion of ambipolar diffusion would change the formation and evolution of the circumstellar disk, 
through the possible contribution in the limited range of density. 

With strong coupling approximation, the induction equation including the ambipolar diffusion can be written as 
\begin{equation}
\dfrac{\partial \vect{B}}{\partial t} = \nabla \times (\vect{v} \times \vect{B}) + \eta_{\rm OD} \nabla^2 \vect{B} + 
\nabla \times 
\left[ 
 \dfrac{\vect{B}}{4\pi \gamma \rho_n \rho_i} \times \left[ \vect{B} \times (\vect{\nabla} \times \vect{B}) \right]
\right],
\label{eq:ambipolar}
\end{equation}
where $\rho_n$, $\rho_i$ and $\gamma$ are the density of neutral gas and charged particles and a drag coefficient between charged particles and neutrals, respectively (for details, see, e.g., \citealt{shu92}).
The third term of equation~(\ref{eq:ambipolar}) can be regarded as a non-linear diffusion term by the ambipolar diffusion.
When we adopt an approximate relation between 
neutral and charged particle density, $\rho_i = C \rho_n^{1/2}$, 
where $C$ is a constant, 
the effective diffusion coefficient of the ambipolar diffusion ($\eta_{\rm ad}$) can be roughly estimated as 
\begin{equation}
\eta_{\rm ad} = \dfrac{B^2}{4\pi \gamma C \rho^{3/2}}. 
\label{eq:coef}
\end{equation}
In a realistic setting, 
\citet{kunz10} showed that diffusion coefficient of the ambipolar diffusion ($\eta_{\rm ad}$) is 10-100 times larger than that of the Ohmic dissipation ($\eta_{\rm OD}$) in the range of $10^{11}\cm \lesssim n \lesssim 10^{12}\cm$. 
Strictly speaking, to include the effect of the ambipolar diffusion, we should solve equation~(\ref{eq:ambipolar}) together with the equations determining charged particle density. 
Note that 
the realistic evolution of charged particle density sensitively depends on total surface area of dust grains, since the recombination of charged particles on the grain surface is very efficient. 
However, there is a considerable uncertainty in the evolution of dust grain properties, such as the size distribution, in the protostellar collapse process. 
Thus, our limited knowledge of grain properties make our accurate treatment of magnetic field dissipation impracticable. 
This is also the reason why we adopted parametric description of the 
Ohmic diffusion coefficient in our study of magnetic field dissipation
\citep{machida07}. 
In this paper we take the same approach to the effect of ambipolar diffusion. 
To roughly estimate the effect of the ambipolar diffusion, we calculate the evolution of the circumstellar disk with an artificially larger resistivity.
In equation~(\ref{eq:reg}), we replace $\eta_{\rm OD}$ by $c_\eta\, \eta_{\rm OD}$ as follows:
\begin{equation}
 \dfrac{\partial \vect{B}}{\partial t} = 
   \nabla \times (\vect{v} \times \vect{B}) + c_\eta \, \eta_{\rm OD} \nabla^2 \vect{B}.
\label{eq:od2}
\end{equation}
We adopt $c_\eta = 0$ (ideal MHD model), 1 (fiducial model), 10, 100.

\begin{figure}
\includegraphics[width=150mm]{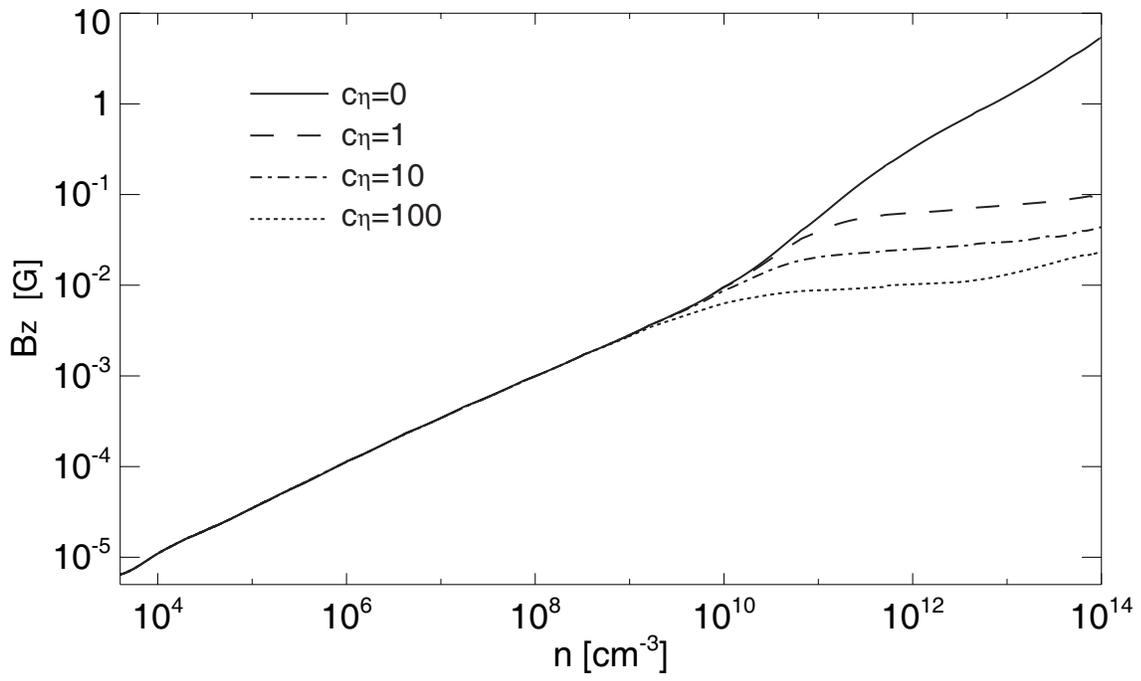}
\caption{
The magnetic flux density against central number density for models with $c_\eta = 0$ (ideal MHD model), 1 , 10 and 100.
}
\label{fig:15}
\end{figure}

As a first test, we constructed a non-rotating Bonner-Ebert sphere with a central density of $n\simeq 10^{3}\cm$ immersed in the uniform magnetic field of $B_{z,0}=5\times10^{-5}$\,G.
Then, we calculated the collapse of this cloud with different $c_\eta$.
Figure~\ref{fig:15} shows the evolution of the magnetic field as a function of the central number density.
The magnetic field in each model tracks the same path in the range of $n\lesssim 10^9\cm$, while the magnetic field for model with $c_\eta = 100$ begins to depart from that for ideal MHD model ($c_\eta=0$) at $n\simeq10^9\cm$.
The figure indicates that non-ideal MHD effect becomes important at the lower density with the larger $c_\eta$.
When the central number density reaches $n=10^{14}\cm$, the magnetic field strengths for model with $c_\eta=0$ and $100$ are $B_z = 7$\,G and $0.01$\,G, respectively.
Thus, the ideal MHD model has about 100 times stronger magnetic field than model with $c_\eta=100$ at this epoch.
At the same epoch, models with $c_\eta=1$ and 10 have $B_z= 0.03$\,G and 0.06\,G of the magnetic field, respectively.
Thus, the difference of the magnetic field strength between $c_\eta=1$ and $100$ (10) is only factor of 6 (2) .

\begin{figure}
\includegraphics[width=150mm]{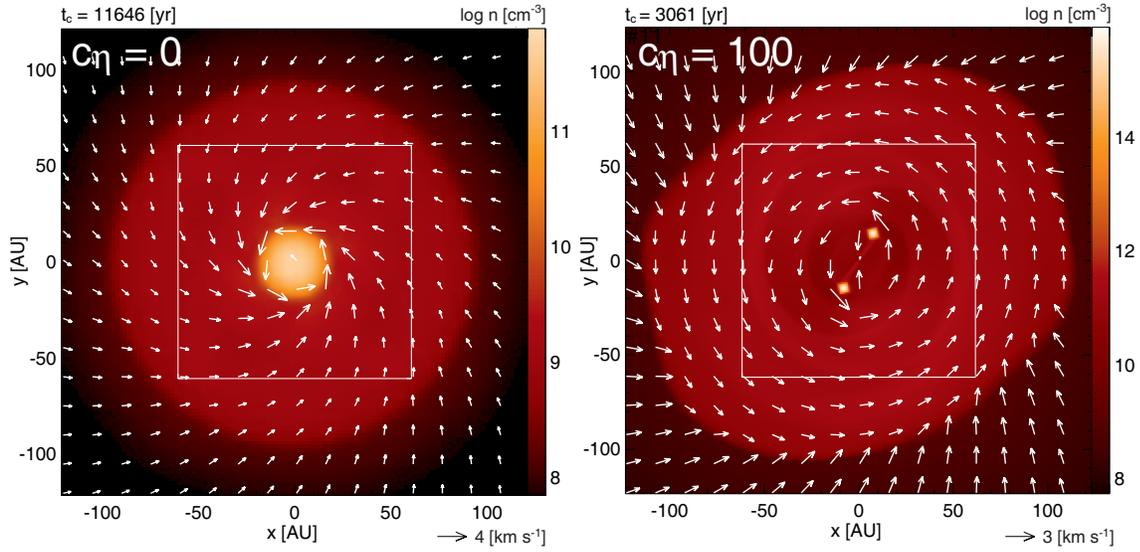}
\caption{
Density distribution and velocity vectors (arrows) on the equatorial plane for models $c_\eta=0$ (ideal MHD; left panel) and 100 (right panel).
The elapsed time after the protostar formation $t_c$ is described in the upper part of each panel.
}
\label{fig:16}
\end{figure}

Next, we calculated the evolution of the circumstellar disk with the initial condition same as model A03 but different resistivities $c_\eta=0$, 10 and 100.
The density distribution on the equatorial plane after the circumstellar disk formation for models with $c_\eta = 0$ and $c_\eta=100$ are plotted in Figure~\ref{fig:16}.
As seen in this figure, fragmentation does not occur in ideal MHD model ($c_\eta=0$) until the end of the calculation, while, for model with $c_\eta=100$, fragmentation occurs at $t_c \simeq 2550$\,yr after the protostar formation and two clumps appear in the circumstellar disk.
For model with $c_\eta=10$, the circumstellar disk also shows fragmentation and subsequent planet formation.
Note that we have terminated the calculation with larger resistivity earlier, since it takes formidable time to follow the evolution with larger resistivity with our time-explicit integration scheme for diffusion term \citep{machida07}. 
Thus, we cannot directly compare the results among models with different resistivities.
However, Figure~\ref{fig:16} indicates that fragmentation occurs even when the magnetic field dissipates in a relatively low-density region.
Therefore we expect that recurrent planet formation and intermittent outflow can occur even with more realistic treatment of the ambipolar diffusion.

\subsection{Comparison with Previous Studies}
Only a few studies focused on the formation and evolution of the circumstellar disk from the prestellar core stage.
\citet{walch09a} and \citet{vorobyov10} investigated the circumstellar disk in unmagnetized clouds, and found that massive circumstellar disk formed at the early main accretion phase tends to show fragmentation.
\citet{machida10} and \citet{bate10} pointed out that the disk forms before the protostar formation, and the circumstellar disk is inevitably massive than the protostar in the early accretion phase.
This is because the circumstellar disk is originated from the first adiabatic core formed in the gas-collapsing phase before the protostar formation \citep{bate98,machida10}.
Such massive circumstellar disk is prone to show fragmentation.
These studies indicate the possibility of the formation of substellar- or planet-mass object in the unmagnetized circumstellar disk by the gravitational instability.

In the magnetized cloud core, only \citet{vorobyov05,vorobyov06} investigated the formation and evolution of the circumstellar disk using two-dimensional simulation.
Although, they adopted the magnetic evolution in an approximate way instead of solving the induction equation, they presented several interesting results.
They pointed out that fragmentation and planet formation is possible even in the magnetized circumstellar disk.
They also showed that the gravitational instability or fragments temporarily amplifies the mass accretion rate onto the protostar that may account for the FU Orionis outbursts.
There are slight quantitative differences between \citet{vorobyov05,vorobyov06} and ours, because we calculated the circumstellar disk with a realistic process of the magnetic dissipation in three dimensions.
However, our results are qualitatively the same as their results.
This is because the circumstellar disk always experiences gravitationally unstable phase after the protostar formation.
\citet{inutsuka09} showed that, even in the magnetized cloud core, the circumstellar disk is massive than the protostar and becomes gravitationally unstable.
Such disk tends to show fragmentation in the early accretion phase.
Even when fragmentation does not occur in the early accretion phase, such massive disk is expected to show fragmentation in the later accretion  or subsequent phase.
Thus, we expect that the planet formation due to the gravitational instability is common in the star formation process.

\section{Summary}
To investigate the properties of the circumstellar disk during the main accretion phase, we calculated the formation and evolution of the circumstellar disk and protostellar outflow in a magnetized collapsing cloud core from the prestellar core stage using three-dimensional nested-grid simulation that covers both the whole region of a nascent molecular cloud core and sub-AU structure.
We constructed two models with a parameter for the initial stability of cloud core  (the ratio of thermal energy to gravitational energy) or total mass of the cloud core, in which the rotation rate and magnetic field strength were fixed, and calculated the evolution of the circumstellar disk about $\sim10^4$\,yr after protostar formation.
The different initial stabilities affect the rate of mass accretion onto the circumstellar disk; that is, a relatively massive disk appears with a high mass accretion rate in a relatively thermally unstable cloud core.
We obtained the following results from these calculations:

\begin{itemize}
\item {\it \bf Recurrent Planet Formation in the Circumstellar Disk}\\
The circumstellar disk formed in a relatively unstable cloud core is considerably massive because of the high rate of mass accretion onto the disk and is in a highly gravitationally unstable state. 
Thus, fragmentation occurs and  substellar-mass objects appear in the circumstellar disk.
The substellar-mass objects lose their orbital angular momenta by gravitational interaction with the massive circumstellar disk and finally fall onto the protostar for $\sim1000$\,yr after their formation.
After those objects fall, the circumstellar disk increases its mass by the mass accretion from the nascent cloud core and again shows fragmentation and subsequent falling of fragments onto the protostar.

Fragmentation and  falling are repeated in the main accretion phase.
The mass of fragments never reaches the hydrogen-burning limit of $\sim0.08\msun$ because the circumstellar disk self-regulates its mass during this recurrent planet formation process, that is,  the circumstellar disk induces fragmentation and loses its mass when fragments fall onto the protostar before it becomes vary massive.
Thus, fragments formed in the main accretion phase hardly exceed the planet or brown-dwarf mass.
It is expected that this recurrent planet formation process continues as long as the disk accretes mass at the high rate.
However, mass accretion onto the disk decreases with time.
Thus, the final generation of fragments formed in the outer disk region just before the mass accretion stops (or significantly decreases) may survive to become long-lived planets or brown-dwarfs with long orbital periods, as recently observed \citep{kalas08,marois08,thalmann09}.

\item {\it \bf Highly Time-Variable Mass Accretion}\\
In the main accretion phase, both models (relatively stable and unstable models) show non-steady mass accretion onto the central protostar.
In a relatively stable cloud core that shows no fragmentation, the rate of mass accretion onto the protostar fluctuates in the range of $\sim10^{-6}$ to $\sim10^{-4}\,\mdot$ with a period of $\sim100$\,yr.
Thus,  highly time-variable mass accretion is seen in this model.
The magnitude of the mass accretion rate almost corresponds to the theoretical prediction.
The cycle of variation of the mass accretion rate (or the accretion luminosity) corresponds to the orbital period of the location of clear spiral arms that effectively transfer the angular momentum outward and promote protostellar mass accretion.
The spiral arms only appear in the inner disk region where a considerably weak field is realized owing to the effective Ohmic dissipation, and the angular momentum is not effectively transferred by the magnetic effect.
As a result, the excess angular momentum causes the mass to accumulate in the inner disk region and increases the non-axisymmetric perturbation that promotes the mass accretion onto the protostar.

Instead, in the outer disk region where the magnetic field is well-coupled with neutrals, the non-asymmetric perturbation hardly grows because the angular momentum is effectively transferred by the magnetic effect.
In summary, because the spiral structure induces the time-variable protostellar accretion, the typical period of the variation of the accretion luminosity corresponds to the orbital extent of the magnetically inactive zone where the spiral patterns appear.
Thus, the time variation or the cycle of the accretion luminosity is related to the magnetic dissipation process that is connected to the magnetic flux problem.
Although we adopted a very realistic value of the magnetic resistivity and ionization degree, we may also deduce them by observations of the luminosity variation of Class 0 and I objects.
In addition, the time variation of the accretion luminosity caused by spiral patterns may explain the luminosity dispersion of Class 0 and I objects 
over 2-3 orders of magnitude \citep{evans09}.

\item {\it \bf Episodic Mass Accretion}\\
Episodic mass accretion is seen in a relatively unstable cloud core.
Fragmentation occurs and substellar-mass objects repeatedly appear in the circumstellar disk formed in an unstable cloud core.
The substellar-mass objects disturb the circumstellar disk and induce a strong episodic brightening of the protostar.
The mass accretion rate  is considerably low $\lesssim 10^{-6}\,\mdot$\  when the substellar-mass objects orbit in a region sufficiently far ($\gg5$\,AU) from the protostar, while the rate  increases to $\gtrsim 10^{-5}\,\msun$  as the substellar-mass objects approach ($\lesssim5$\,AU) the protostar.
Thus, during the main accretion phase, the protostar shows two distinct phases of mass accretion: quiescent and active mass accretion phases, as seen in \citet{vorobyov06}.
The protostar increases its luminosity ($\gtrsim1\lsun$) by a significant mass accretion in the active phase that lasts only for $\sim100$\,yr, while it dims ($\lesssim 0.1\lsun$) in the quiescent phase that lasts for $\gtrsim1000$\,yr.
Recent observations support episodic mass accretion to solve the luminosity problem \citep{enoch09b,dunham10}.
In addition, FUori-type outbursts also may be related to the episodic accretion caused by orbital motion and falling of the substellar-mass objects.
However, we need further spatial resolution to associate them with actual observations.

\item {\it \bf Intermittent Protostellar Outflows}\\
Both models show the outflow driven by the circumstellar disk.
The steady outflow that lasts for $>10^4$\,yr appears in a relatively stable cloud core, while the non-steady outflow appears in a relatively unstable cloud core.
In a relatively stable cloud core, spiral arms that promote mass accretion onto the protostar appear in the inner disk region where magnetic dissipation is effective and no outflow appears, while the  outflow is driven by the outer disk region where an almost axisymmetric structure remains.
As a result, time-variable mass accretion is caused by the inner disk region, while outflow is steadily driven by the stable outer disk region.

On the other hand, in a relatively unstable cloud core, the outflow driven by the circumstellar disk shows a strong time variability because substellar-mass objects disturb magnetic field lines in the whole circumstellar disk.
As a result, the episodic (or intermittent) outflow appears with episodic mass accretion.
Thus, intermittent outflow may be proof of the existence of planets or substellar-mass objects in the circumstellar disk.
In addition, variable periods of the outflow well synchronize with the orbital period of substellar-mass objects. 
Thus, we may be able to predict the existence of a planet and its orbital period with detailed observation of the trail of the protostellar outflow.
However, further long-term calculations are necessary to investigate outflow propagation in the interstellar medium.
\end{itemize}

\acknowledgments
Numerical computations were carried out on NEC SX-9 at Center for Computational Astrophysics, CfCA, of National Astronomical Observatory of Japan, and NEC SX-8 at the Yukawa Institute Computer Facility.
This work was supported by Grants-in-Aid from MEXT (20540238, 21740136).

\clearpage
\begin{table}
\setlength{\tabcolsep}{3pt}
\caption{Model parameters and calculation results}
\label{table}
\footnotesize
\begin{center}
\begin{tabular}{c|ccccccc|cccccc} \hline
{\footnotesize Model} & $\alpha_0$ &  $f$ & $\Omega_0$ {\scriptsize [s$^{-1}$]} & $B_{\rm ini}$  {\scriptsize [$\mu$G]}
& $M_{\rm cl}$ {\scriptsize [$\msun$]} & $R_{\rm c}$ {\scriptsize [AU]} & $\lambda$
 
& $M_{\rm ps}$ {\scriptsize [$\msun$]} & $M_{\rm disk}$ {\scriptsize [$\msun$]} &  N$_{\rm pl}$ \\ \hline
A03     & 0.3 & 2.8 & $1.1\times10^{-13}$ & 37 & 1.3 & 4700 & 9 & 0.59 & 0.15 & 16 \\
A05     & 0.5 & 1.7 & $1.1\times10^{-13}$ & 37 & 0.8 & 4700 & 5.6 &0.15 & 0.03 &  0 \\
\hline
\end{tabular}
\end{center}
\end{table}

\begin{table}
\caption{Properties of planets}
\label{table}
\footnotesize
\begin{center}
\begin{tabular}{c|cccccccccccc} \hline
Generation & $r_i$ {\scriptsize [AU]} &  $t_{\rm i}$ {\scriptsize [yr]} & $t_{\rm e}$ {\scriptsize [yr]} & $t_{e}-t_{i}$ {\scriptsize [yr]} & $M_{\rm pl}$ {\scriptsize $[M_{\rm Jup}]$} &  \\ \hline
1 & 6.8 & 795 & 3657 & 2862 & 23 \\
2 & 45.8 & 3990 & 4669 & 679 & 6 \\
3 & 22.5 & 5037 & 5693 & 656 & 26 \\
4 & 36.9 & 5984 & 6603 & 619 &16 \\
5 & 36.5 & 6697 & 7503 & 806 &15 \\
6 & 40.2 & 7504 & 8446 & 942 & 9 \\
7 & 23.3 & 8455 & 9240 & 785 & 27 \\
8 & 25.9 & 9327 & 9760 & 433 & 25 \\
9 & 22.1 & 9748 & --- & ---  &12 \\
\hline
\end{tabular}
\end{center}
\end{table}

\clearpage
\clearpage

\clearpage

\clearpage

\clearpage

\clearpage

\clearpage

\clearpage

\clearpage

\clearpage

\clearpage

\clearpage

\clearpage

\clearpage

\clearpage


\begin{thebibliography}{}{}
\bibitem[Bate(1998)]{bate98} 
 Bate, M.~R.\ 1998, \apjl, 508, L95

\bibitem[Bate(2010)]{bate10} 
 Bate, M.~R.\ 2010, \mnras, L38

\bibitem[\protect\citeauthoryear{Bate et al.}{2003}]{bate03} 
Bate, M.~R., Lubow, S.~H., Ogilvie, G.~I., \& Miller, K.~A.\ 2003, MNRAS, 341, 213 


\bibitem[Banerjee \& Pudritz(2006)]{banerjee06}
 Banerjee, R., \& Pudritz, R. E. 2006, \apj, 641, 949

\bibitem[Bodenheimer \etal(2000)]{bodenheimer00}
 Bodenheimer P., Burkert A., Klein R. I., \& Boss A. P., 2000, in Mannings V., Boss A. P., Russell S. S., eds, Protostars and Planets IV. Univ. Arizona Press, , p. 675

\bibitem[Cameron(1978)]{cameron78}
 Cameron, A.~G.~W.\ 1978, Moon and Planets, 18, 5 

\bibitem[Crutcher(1999)]{crutcher99}
 Crutcher R. M. 1999, ApJ, 520, 706

\bibitem[\protect\citeauthoryear{D'Angelo \etal}{2003}]{dangelo03}
 D'Angelo, G., Kley, W., \& Henning, T. 2003, ApJ, 586, 540

\bibitem[Dedner \etal (2002)]{dedner02}
 Dedner A., Kemm F., Kr\"oner D., Munz C.-D., Schnitzer T., Wesenberg M., 2002, J. Comp. Phys., 175, 645

\bibitem[Durisen et al.(2007)]{durisen07} 
Durisen, R.~H., Boss, A.~P., Mayer, L., Nelson, A.~F., Quinn, T., \& Rice, W.~K.~M.\ 2007, Protostars and Planets V, 607 
\bibitem[Dunham et al.(2010)]{dunham10} 
Dunham, M.~M., Evans, N.~J., Terebey, S., Dullemond, C.~P., \& Young, C.~H.\ 2010, \apj, 710, 470 

\bibitem[Enoch et al.(2009a)]{enoch09a} 
Enoch, M.~L., Corder, S., Dunham, M.~M., \& Duch{\^e}ne, G.\ 2009a, \apj, 707, 103 

\bibitem[Enoch et al.(2009b)]{enoch09b} 
Enoch, M.~L., Evans, N.~J., Sargent, A.~I., \& Glenn, J.\ 2009b, \apj, 692, 973 

\bibitem[Evans et al.(2009)]{evans09} 
Evans, N.~J., et al.\ 2009, \apjs, 181, 321

\bibitem[Goodwin \etal(2007)]{goodwin07}
 Goodwin S. P., Kroupa P., Goodman A., \& Burkert A., 2007, in Reipurth B., Jewitt D., Keil K., eds, Protostars and Planets V. Univ. Arizona Press, , p. 133

\bibitem[Hayashi \etal(1985)]{hayashi85}
 Hayashi, C., Nakazawa, K., \& Nakagawa, Y. 1985, in Protostars and Planets II, ed. D. C. Black \& M. S. Matthews (Tucson: Univ. Arizona Press), 110

\bibitem[Hennebelle \& Fromang(2008a)]{hennebelle08a} 
Hennebelle, P., \& Fromang, S.\ 2008a, \aap, 477, 9 

\bibitem[Hennebelle \& Teyssier(2008b)]{hennebelle08b}
 Hennebelle, P., \& Teyssier, R.\ 2008b, \aap, 477, 25

\bibitem[Hunter(1977)]{hunter77}
 Hunter, C.\ 1977, \apj, 218, 834 

\bibitem[Inutsuka et al.(2009)]{inutsuka09} 
Inutsuka, S., Machida, M.~N., \& Matsumoto, T.\ 2009, ApJ accepted (arXiv:0912.5439) 

\bibitem[Kalas et al.(2008)]{kalas08} 
 Kalas, P., et al.\ 2008, Science, 322, 1345 

\bibitem[Kenyon et al.(1990)]{kenyon90} 
Kenyon, S.~J., Hartmann, L.~W., Strom, K.~M., \& Strom, S.~E.\ 1990, \aj, 99, 869 

\bibitem[\protect\citeauthoryear{Kley \etal}{2001}]{kley01}
 Kley, W., D'Angelo, G., \& Henning, T. 2001, ApJ, 547, 457

\bibitem[Kunz \& Mouschovias(2010)]{kunz10} 
 Kunz, M.~W., \& Mouschovias, T.~C.\ 2010, \mnras, 408, 322 

\bibitem[Larson(1969)]{larson69} 
 Larson, R. B., 1969, MNRAS, 145, 271.

\bibitem[Machida et al.(2004)]{machida04} 
Machida, M.~ N., Tomisaka, K., \& Matsumoto, T.\ 2004, \mnras, 348, L1 

\bibitem[Machida \etal(2005a)]{machida05a}
 Machida, M. N., Matsumoto, T., Tomisaka, K., \& Hanawa, T. 2005, MNRAS, 362, 369

\bibitem[Machida \etal (2005b)]{machida05b} 
 Machida, M. N., Matsumoto, T., Hanawa, T., \& Tomisaka, K. 2005b, MNRAS, 362, 382 

\bibitem[Machida et al.(2006)]{machida06} 
Machida, M.~N., Inutsuka, S., \& Matsumoto, T.\ 2006, \apjl, 647, L151 

\bibitem[Machida et al.(2007)]{machida07} 
 Machida, M.~N., Inutsuka, S., \& Matsumoto, T.\ 2007, \apj, 670, 1198 

\bibitem[Machida et al.(2008a)]{machida08a} 
 Machida, M.~N., Tomisaka, K., Matsumoto, T., \& Inutsuka, S. \ 2008a, \apj, 677, 327 

\bibitem[Machida et al.(2008b)]{machida08b} 
Machida, M.~N., Inutsuka, S., \& Matsumoto, T.\ 2008b, \apj, 676, 1088

\bibitem[Machida et al.(2009a)]{machida09a} 
 Machida, M.~N., Inutsuka, S., \& Matsumoto, T.\ 2009a, \apjl, 699, L157

\bibitem[Machida et al.(2009b)]{machida09b} 
Machida, M.~N., Inutsuka, S., \& Matsumoto, T.\ 2009b, \apjl, 704, L10 

\bibitem[Machida et al.(2010)]{machida10} 
Machida, M.~N., Inutsuka, S., \& Matsumoto, T.\ 2010, arXiv:1001.1404 

\bibitem[Machida et al.(2010b)]{machida10b} 
Machida, M.~N., Kokubo, E., Inutsuka, S., \& Matsumoto, T.\ 2010b, arXiv:1002.3002 

\bibitem[Matsumoto \& Hanawa(2003)]{matsu03} 
 Matsumoto T.,  \& Hanawa T., 2003, \apj, 595, 913

\bibitem[Matsumoto \& Tomisaka(2004)]{matsu04}
 Matsumoto, T., \& Tomisaka, K.\ 2004, \apj, 616, 266 

\bibitem[Matsumoto(2007)]{matsu07} 
Matsumoto, T.\ 2007, \pasj, 59, 905 

\bibitem[Marois et al.(2008)]{marois08} 
Marois, C., Macintosh, B., Barman, T., Zuckerman, B., Song, I., Patience, J., Lafreni{\`e}re, D., 
\& Doyon, R.\ 2008, Science, 322, 1348 

\bibitem[Masunaga \& Inutsuka (2000)]{masunaga00} 
 Masunaga, H., \& Inutsuka, S., 2000, ApJ, 531, 350

\bibitem[Mouschovias \& Spitzer(1976)]{mouschovias76}
 Mouschovias, T. Ch., \& Spitzer, L. 1976, \apj, 210, 326

\bibitem[Nakano et al.(2002)]{nakano02} 
Nakano, T., Nishi, R., \& Umebayashi, T.\ 2002, \apj, 573, 199 

\bibitem[Saigo \& Tomisaka(2006)]{saigo06}
 Saigo, K., \& Tomisaka, K.\ 2006, \apj, 645, 381 

\bibitem[Stamatellos et al.(2007)]{stamatellos07} 
 Stamatellos, D., Whitworth, A.~P., Bisbas, T., \& Goodwin, S.\ 2007, \aap, 475, 37 

\bibitem[Shu(1977)]{shu77} 
 Shu, F.~H.\ 1977, \apj, 214, 488 

\bibitem[Shu(1983)]{shu83} 
 Shu, F.~H.\ 1983, \apj, 273, 202 

\bibitem[Shu(1992)]{shu92} 
Shu, F.~H.\ 1992, Physics of Astrophysics, Vol.~II, by Frank H.~Shu.~Published by University Science Books, ISBN 0-935702-65-2, 476pp, 1992.,  

\bibitem[Tassis \& Mouschovias(2007a)]{tassis07a} 
 Tassis, K., \& Mouschovias, T.~C.\ 2007a, \apj, 660, 370 

\bibitem[Tassis \& Mouschovias(2007b)]{tassis07b} 
 Tassis, K., \& Mouschovias, T.~C.\ 2007b, \apj, 660, 388 

\bibitem[Tassis \& Mouschovias(2007c)]{tassis07c} 
 Tassis, K., \& Mouschovias, T.~C.\ 2007c, \apj, 660, 402 


\bibitem[Thalmann et al.(2009)]{thalmann09} 
Thalmann, C., et al.\ 2009, \apjl, 707, L123

\bibitem[Toomre(1964)]{toomre64} 
 Toomre, A.\ 1964, \apj, 139, 1217 

\bibitem[Tomisaka(2002)]{tomisaka02} 
 Tomisaka K., 2002, \apj, 575, 306

\bibitem[Truelove \etal(1997)]{truelove97}
 Truelove J, K., Klein R. I., McKee C. F., Holliman J. H., Howell L. H., \& Greenough J. A., 1997, ApJ, 489, L179

\bibitem[Vorobyov \& Basu(2005)]{vorobyov05} 
 Vorobyov, E.~I., \& Basu, S.\ 2005, \apjl, 633, L137 

\bibitem[Vorobyov \& Basu(2006)]{vorobyov06} 
Vorobyov, E.~I., \& Basu, S.\ 2006, \apj, 650, 956 

\bibitem[Vorobyov \& Basu(2007)]{vorobyov07} 
 Vorobyov, E.~I., \& Basu, S.\ 2007, \mnras, 381, 1009 

\bibitem[Vorobyov \& Basu(2010)]{vorobyov10} 
 Vorobyov, E.~I., \& Basu, S.\ 2010, arXiv:1003.4315 

\bibitem[Whitehouse \& Bate(2006)]{whitehouse06} 
Whitehouse, S.~C., \& Bate, M.~R.\ 2006, \mnras, 367, 32 

\bibitem[Walch et al.(2009a)]{walch09a} 
Walch, S., Burkert, A., Whitworth, A., Naab, T., \& Gritschneder, M.\ 2009a, \mnras, 400, 13 

\bibitem[Walch et al.(2009b)]{walch09b} 
Walch, S., Naab, T., Whitworth, A., Burkert, A., \& Gritschneder, M.\ 2009, \mnras, 1912 

\end{thebibliography}
\end{document}